\documentclass[10pt,journal,compsoc]{IEEEtran}
\usepackage[nocompress]{cite}

\usepackage{enumitem}
\usepackage{bm}
\usepackage{xspace}
\usepackage{color}
\usepackage{mathtools}
\usepackage{subcaption}
\usepackage{amsfonts} % added for SIAM template
\usepackage{amssymb} % added for SIAM template
\usepackage{booktabs}
\usepackage{multirow}
\usepackage{adjustbox}
\usepackage{hyperref}
\usepackage{color,hyperref}
\definecolor{darkblue}{rgb}{0.0,0.0,0.3}
\hypersetup{colorlinks,breaklinks,linkcolor=darkblue,urlcolor=darkblue,anchorcolor=darkblue,citecolor=darkblue}
\usepackage[algo2e, ruled, vlined]{algorithm2e}
\usepackage[font=small,labelfont=bf,tableposition=top]{caption}

\newcommand{\vpara}[1]{\vspace{0.08in}\noindent\textbf{#1 }}
\newcommand{\hide}[1]{} %hide

\newcommand\hmm[1]{\ifnum\ifhmode\spacefactor\else2000\fi>1000 \uppercase{#1}\else#1\fi}

\newcommand{\ie}{{\sl i.e.\xspace}}

\newcommand{\wrt}{{\sl w.r.t.\xspace}}

\newtheorem{thm}{Theorem}[section] % added for SIAM template
\newtheorem{definition}[thm]{Definition} % added for SIAM template

\newcommand{\mc}{\mathcal}
\newcommand{\expectation}{\mathbb{E}}
\newcommand{\mbf}{\mathbf{f}}
\newcommand{\mbg}{\mathbf{g}}
\newcommand{\norm}[1]{\left\lVert#1\right\rVert_2}
\renewcommand{\phi}{\varphi}

\newcommand{\funcphi}[1]{\phi^v_\theta(\{\tilde\mbg^{v'}_{#1}\}_{v' \in \mc{V}})}

\newcommand{\preservation}{\textit{preservation}\xspace}
\newcommand{\collaboration}{\textit{collaboration}\xspace}
\newcommand{\independent}{\textit{independent}\xspace}
\newcommand{\onespace}{\textit{one-space}\xspace}
\newcommand{\singleview}{\textit{single-view}\xspace}
\newcommand{\viewmerging}{\textit{view-merging}\xspace}
\newcommand{\mtv}{\textsc{mvn2vec}\xspace}
\newcommand{\mtvreg}{\textsc{mvn2vec-reg}\xspace}
\newcommand{\mtvcon}{\textsc{mvn2vec-con}\xspace}

\begin{document}
\title{\mtv{}: Preservation and Collaboration in Multi-View Network Embedding}
\author{Yu~Shi,~\IEEEmembership{Member,~IEEE},
        Fangqiu~Han,
        Xinwei~He,
        Xinran~He,
        Carl~Yang,~\IEEEmembership{Member,~IEEE},
        Roger~Jie~Luo,
        and~Jiawei~Han,~\IEEEmembership{Fellow,~IEEE}
\IEEEcompsocitemizethanks{\IEEEcompsocthanksitem 
Y. Shi, Xinwei He, C. Yang, and J. Han are with University of Illinois at Urbana Champaign, Urbana, IL 61801.
F. Han, Xinran He, and R. J. Luo are with Snap Inc., Santa Monica, CA 90405.
Email: \{yushi2, xhe17, jiyang3, hanj\}@illinois.edu, \{fangqiu.han, xhe2, roger.luo\}@snap.com.
This research was partially conducted when the first author interned at Snap Research.
\protect}
%\IEEEcompsocthanksitem J. Doe and J. Doe are with Anonymous University.}% <-this % stops an unwanted space
%\thanks{Manuscript received --; revised --.}
}

\markboth{}
{Shi \MakeLowercase{\textit{et al.}}: \mtv{}: Preservation and Collaboration in Multi-View Network Embedding}

\IEEEtitleabstractindextext{
%!TEX root = mvn2vec_siam.tex

\begin{abstract}
Multi-view networks are broadly present in real-world applications.
In the meantime, network embedding has emerged as an effective representation learning approach for networked data.
Therefore, we are motivated to study the problem of multi-view network embedding with a focus on the optimization objectives that are specific and important in embedding this type of network.
In our practice of embedding real-world multi-view networks, we explicitly identify two such objectives, which we refer to as \preservation and \collaboration.
The in-depth analysis of these two objectives is discussed throughout this paper.
In addition, the \mtv algorithms are proposed to (i) study how varied extent of \preservation and \collaboration can impact embedding learning and (ii) explore the feasibility of achieving better embedding quality by modeling them simultaneously.
With experiments on a series of synthetic datasets, a large-scale internal Snapchat dataset, and two public datasets, we confirm the validity and importance of \preservation and \collaboration as two objectives for multi-view network embedding.
These experiments further demonstrate that better embedding can be obtained by simultaneously modeling the two objectives, while not over-complicating the model or requiring additional supervision.
The code and the processed datasets are available at \url{http://yushi2.web.engr.illinois.edu/}.
\end{abstract}

}

\maketitle
\IEEEdisplaynontitleabstractindextext
\IEEEpeerreviewmaketitle

%\IEEEraisesectionheading{\section{Introduction}\label{sec:introduction}}
%\IEEEPARstart{T}{his} demo file is intended to serve as a ``starter file''
%\hfill mds
%\hfill August 26, 2015

%!TEX root = mvn2vec_siam.tex

\section{Introduction}\label{sec::introduction}

\begin{figure}[t!]
    \begin{subfigure}[t]{0.225\textwidth}
        \centering
        \includegraphics[width=\textwidth]{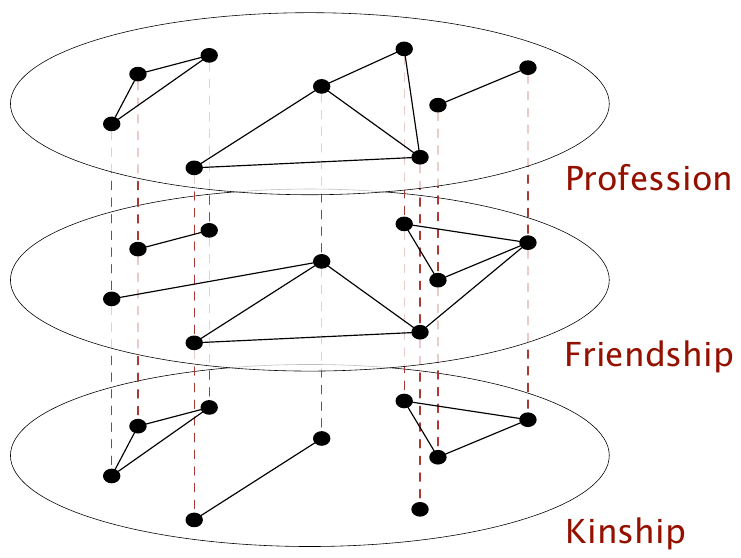}
        \caption{A toy multi-view network.}\label{fig::views-together}
    \end{subfigure}
    ~
    \hspace*{0.01\textwidth}
    \begin{subfigure}[t]{0.225\textwidth}
        \centering
        \includegraphics[width=\textwidth]{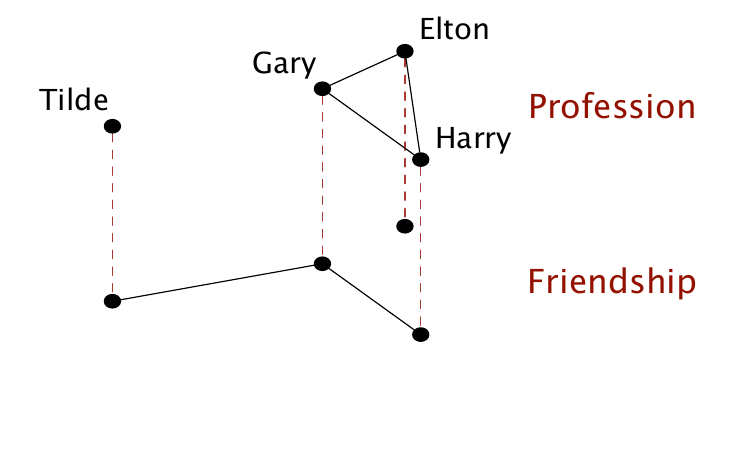}
        \caption{A close-up look of profession view and friendship view.}\label{fig::close-up}
    \end{subfigure}    
%\vspace{-3pt}
\caption{A toy example of multi-view networks where each node represents a person and the three views correspond to three types of interpersonal relations. Co-workers are linked in the profession view, friends are linked in the friendship view, and relatives are linked in the kinship view.}
\end{figure}

In real-world applications, objects can be associated with different types of relations.
These objects and their relationships can be naturally represented by multi-view networks, \ie, multiplex networks or multi-view graphs~\cite{kumar2011co-r, zhou2007spectral, sindhwani2005co, pei2005mining, frank1993exploratory, pattison1999logit}.
Figure~\ref{fig::views-together} gives a toy multi-view network, where each view corresponds to a type of edge, and all views share the same set of nodes.
%In ecology, a multi-view network can be used to represent the relation among species, where each node stands for a species, and the six views represent predation, competition, symbiosis, parasitism, protocooperation, and commensalism, respectively.
As a more concrete example, a four-view network of users can be used to describe a social networking service with social relationship and interaction including friendship, following, message exchange, and post viewing.
With the vast availability of multi-view networks, it is of interest to mine such networks. %, such as clustering species in the multi-view ecological networks, or improving product design by mining the multi-view social networks.
%In order to achieve this goal with the progressively developed computing power, it is of interest to first transform the multi-view networks into a different form of representations that are more machine actionable.

In the meantime, network embedding has emerged as a scalable representation learning method for networked data ~\cite{grover2016node2vec, perozzi2014deepwalk, tang2015line, wang2016structural}.
Specifically, network embedding projects nodes of networks into the embedding spaces.
With the semantic information of each node encoded, the learned embedding can be directly used as features in various downstream applications~\cite{grover2016node2vec, perozzi2014deepwalk, tang2015line}.
Motivated by the success of network embedding for homogeneous networks~\cite{grover2016node2vec, perozzi2014deepwalk, tang2015line, wang2016structural, perozzi2017don, ou2016asymmetric}, where nodes and edges are untyped, we believe it is important to better understand multi-view network embedding.

To design embedding method for multi-view networks, the primary challenge lies in how to use the type information on edges from different views.
As a result, we are interested in investigating the following two questions: 
%\vspace{-3pt}
\begin{enumerate}
\item
With the availability of multiple edge types, what are the objectives that are \textit{specific} and \textit{important} to multi-view network embedding?
%\vspace{-3pt}
\item
Can we achieve better embedding quality by modeling these objectives jointly?
%\vspace{-3pt}
\end{enumerate}

To answer the first question, we identify two such objectives, \preservation and \collaboration, from our practice of embedding real-world multi-view networks.
%We describe the concepts of \preservation and \collaboration as follows.
\textbf{Collaboration} -- In some datasets, edges between the same pair of nodes may be observed in different views due to shared latent reasons.
For instance, in a social network, if we observe an edge between a user pair in either the message exchange view or the post viewing view, likely these two users are happy to be associated with each other.
In such scenario, these views may complement each other, and embedding them jointly may potentially yield better results than embedding them independently.
We call such synergetic effect in jointly embedding multiple views by \collaboration, which is also the primary intuition behind most existing multi-view network algorithms~\cite{kumar2011co-r, zhou2007spectral, sindhwani2005co, pei2005mining}.
\textbf{Preservation} -- On the other hand, it is possible for different network views to have different semantic meanings; it is also possible that a portion of nodes has completely disagreeing edges in different views since edges in different views are formed due to unrelated latent reasons.
For example, the professional relationship may not always align well with friendship.
If we embed the profession view and the friendship view in Figure~\ref{fig::close-up} into the same embedding space, the embedding of Gary will be close to both Tilde and Elton.
As a result, the embedding of Tilde will also not be too distant from Elton due to transitivity.
However, this is not a desirable result, because Tilde and Elton are not closely related regarding either profession or friendship according to the original multi-view network.
In other words, embedding in this way fails to preserve the unique information carried by different network views.
We refer to such need for preserving unique information carried by different views as \preservation.
The detailed discussion of the validity and importance of \preservation and \collaboration is presented in Section~\ref{sec::intuition}.

Furthermore, the need for \preservation and \collaboration may co-exist in the same multi-view network.
Two scenarios can result in this situation: (i) a pair of views are generated from very similar latent reason, while another pair of views carries completely different semantic meanings; and more subtly (ii) for the same pair of views, one portion of nodes has consistent edges in different views, while another portion of nodes have totally disagreeing edges in different views.
One example of the latter scenario is that professional relationship does not align well with friendship in some cultures, whereas co-workers often become friends in certain other cultures \cite{alston1989wa}. 
Therefore, we are also interested in exploring the feasibility of achieving better embedding quality by modeling \preservation and \collaboration simultaneously, and we address this problem in Section~\ref{sec::modeling} and beyond.

We note that, instead of proposing a sophisticated model that beats many baselines, this paper focus on the objectives of interests for multi-view network embedding. 
In experiments, we compare methods with a highlight on the roles the two objectives play in different scenarios. 
For the same reason, the scenarios where additional supervision is available are excluded from this paper, while the lessons learned from the unsupervised scenario can also be applied to the supervised multi-view network embedding algorithms.
Additionally, node embedding learned by an unsupervised approach can directly apply to different downstream tasks, while supervised algorithms yield embedding specifically good for tasks where the supervision comes from.
We summarize our contributions as follows.
\begin{enumerate}
\item
We study the objectives that are specific and important to multi-view network embedding and identify \preservation and \collaboration as two such objectives from the practice of embedding real-world multi-view networks.
\item
We explore the feasibility of attaining better embedding by simultaneously modeling \preservation and \collaboration, and propose two multi-view network embedding methods -- \mtvcon and \mtvreg.
\item
We conduct experiments with various downstream applications on four datasets.
These experiments corroborate the validity and importance of \preservation and \collaboration and demonstrate the effectiveness of the proposed methods.
\end{enumerate}

%!TEX root = mvn2vec_siam.tex

\section{Related Work}\label{sec::related-work}
\vpara{Network embedding.}
Network embedding has emerged as an efficient and effective approach for learning distributed node representations.
Instead of leveraging spectral properties of networks as commonly seen in traditional unsupervised feature learning approaches~\cite{belkin2001laplacian, roweis2000nonlinear, tenenbaum2000global, yan2007graph}, most network embedding methods are designed atop local properties of networks that involve links and proximity among nodes~\cite{grover2016node2vec, perozzi2014deepwalk, tang2015line, wang2016structural, perozzi2017don, ou2016asymmetric, zhang2017weisfeiler, ribeiro2017struc2vec}.
Such methodology focusing on local properties has been shown to be more scalable. 
The designs of many recent network embedding algorithms are based on local properties of networks and can trace to the skip-gram model~\cite{mikolov2013distributed}, which are shown to be scalable and widely applicable. 
To leverage the skip-gram model, various strategies have been proposed to define the context of a node in the network scenario~\cite{grover2016node2vec, perozzi2014deepwalk, tang2015line, perozzi2017don}.
Beyond the skip-gram model, embedding methods for preserving certain other network properties can also be found in the literature~\cite{wang2016structural, niepert2016learning, velivckovic2017graph, kipf2016semi}.

\vpara{Multi-view networks.}
%Meanwhile, multi-view networks have been extensively studied as a special type of networks.%, motivated by their ubiquitous presence in real-world applications.
The majority of existing methods for multi-view networks aim to bring performance boost in traditional tasks, 
such as clustering \cite{kumar2011co-r, zhou2007spectral}, 
classification \cite{sindhwani2005co}, 
and dense subgraph mining \cite{pei2005mining}.
%The above methods aim to improve the performance of specific applications but do not directly study network embedding distributed representation learning for multi-view networks.
Another line of research focuses on measuring and analyzing cross-view interrelations in multi-view networks~\cite{battiston2014structural, de2013mathematical, berlingerio2013multidimensional, pattison1999logit}, but they do not discuss the objectives of embedding multi-view networks, nor do they study how their proposed measures and analyses can relate to the embedding learning of multi-view networks.

\vpara{Multi-view network embedding.}
An attention-based collaboration framework is recently proposed for multi-view network embedding~\cite{qu2017attention}.
The problem setting of this work differs from ours since it requires supervision for its attention mechanism.
Besides, this approach does not directly model \preservation{} -- one of the objectives that we deem important for multi-view network embedding. 
Since the final embedding derived via linear combination in this framework is a trade-off between representations from all views.
A deep learning architecture has also been proposed for embedding multi-networks~\cite{ni2018co}, where the multi-network is a more general concept than the multi-view network and allows many-to-many correspondence across networks.
While the proposed model can be applied to the more specific multi-view networks, it does not focus on the study of the objectives of multi-view network embedding.
Another group of related work studies the problem of jointly modeling multiple network views using latent space models \cite{gollini2014joint, greene2013producing}.
These works again do not model \preservation.
There exist a few more studies that touches the topic of multi-view network embedding\cite{ma2017multi, zhang2018scalable, liu2017principled, ning2018representation, cen2019representation}
They do not model \preservation and \collaboration or do not aim to provide in-depth study of these objectives.

Besides, the problem of multi-view matrix factorization \cite{greene2009matrix, singh2008relational, liu2013multi} is also related to multi-view network embedding. 
Liu et al. \cite{liu2013multi} propose a multi-view nonnegative matrix factorization model, which aims to minimize the distance between the adjacent matrix of each view and the consensus matrix. 
While these methods can generate a vector for each node; these vectors are modeled to suit certain tasks such as matrix completion or clustering, instead of to serve as distributed representations of nodes that can be used in different downstream applications.

\vpara{Embedding techniques for other typed networks.}
In a bigger scope, a couple of studies have recently been conducted in embedding more general networks, such as heterogeneous information networks (HINs)~\cite{tang2015pte, chang2015heterogeneous, fu2017hin2vec, dong2017metapath2vec}. 
Some of them build the algorithms on top of meta-paths~\cite{fu2017hin2vec, dong2017metapath2vec}, but meta-paths are usually more meaningful for HINs with multiple node types than multi-view networks.
Some other methods are designed for networks with additional side information~\cite{chang2015heterogeneous, dai2016discriminative} or particular structures~\cite{tang2015pte, gui2016large}, which do not apply to multi-view networks.
Most importantly, these methods are designed for networks they intend to embed, and therefore do not focus on the study of the particular needs and objectives for embedding multi-view networks.

Multi-relational network embedding or knowledge graph embedding embodies another direction of research that is relevant to multi-view network embedding~\cite{li2017structural, bordes2013translating, wang2014knowledge, lin2015learning}.
The study on this direction has a focus different and does not directly apply to our scenario because the relationships involved in these networks are often directed in nature. 
A popular family of such methods are translation based, where, in their language, the representation of a head entity $\mathbf{h}$ is associated to that of a tail entity $\mathbf{t}$ by that of a relation $\mathbf{r}$ such that $\mathbf{h} + \mathbf{r}$ is assumed to be similar to $\mathbf{h}$~\cite{li2017structural, bordes2013translating, wang2014knowledge, lin2015learning}.
Such assumption is invalid for multi-view networks with undirected network views such as the friendship view in a social network.

%By projecting all network views onto a low-dimensional latent space, this approach can capture associations across different views.
%Besides, the problem of multi-view matrix factorization \cite{greene2009matrix, singh2008relational, liu2013multi} is also related to multi-view network embedding. 
%Liu et al. \cite{liu2013multi} propose a multi-view nonnegative matrix factorization model, which aims to minimize the distance between the adjacent matrix of each view and the consensus matrix. 
%While these methods can generate a vector for each node; these vectors are modeled to suit certain tasks such as matrix completion or clustering, instead of to serve as distributed representations of nodes that can be used in different downstream applications.

%\cite{abu2017learning}

%!TEX root = mvn2vec_siam.tex

\section{Preliminaries}\label{sec::prelim}

\begin{definition}[Multi-View Network]
A multi-view network $G = (\mc{U}, \{\mc{E}^{(v)}\}_{v \in \mc{V}})$ is a network consisting of a set $\mc{U}$ of nodes and a set $\mc{V}$ of views, where $\mc{E}^{(v)}$ consists of all edges in view $v \in \mc{V}$. If a multi-view network is weighted, then there exists a weight mapping $w: \{\mc{E}^{(v)}\}_{v \in \mc{V}} \rightarrow \mathbb{R}$ such that $w^{(v)}_{u u'} \coloneqq w(e^{(v)}_{u u'})$ is the weight of the edge $e^{(v)}_{u u'} \in \mc{E}^{(v)}$, which joints nodes $u \in \mc{U}$ and $u' \in \mc{U}$ in view $v \in \mc{V}$.
\end{definition}

Additionally, when context is clear, we use the network view $v$ of $G = (\mc{U}, \{\mc{E}^{(v)}\}_{v \in \mc{V}})$ to denote the network $G^{(v)} = (\mc{U}, \mc{E}^{(v)})$.

\begin{definition}[Network Embedding]
Network embedding aims at learning a (center) embedding $\mbf_u \in \mathbb{R}^D$ for each node $u \in \mc{U}$ in a network, where $D \in \mathbb{N}$ is the dimension of the embedding space.
\end{definition}
%\vspace{-3pt}

Besides the center embedding $\mbf_u \in \mathbb{R}^D$, a family of popular algorithms~\cite{mikolov2013distributed, tang2015line} also deploy a context embedding $\tilde\mbf_u \in \mathbb{R}^D$ for each node $u$.
Moreover, when the learned embedding is used as the feature vector for downstream applications, we take the center embedding of each node as feature following the common practice in algorithms involving context embedding.

\begin{table}[t!]
\centering
\caption{Summary of symbols}\label{tab::symbol}
%\vspace{-9pt}
\resizebox{0.48\textwidth}{!}{
\begin{tabular}{| c | c |}
\hline
\textbf{Symbol}  & \textbf{Definition} \\ 
\hline \hline
$\mc{V}$ & The set of all network views\\
$\mc{U}$ & The set of all nodes\\
$\mc{E}^{(v)}$ & The set of all edges in view $v \in \mc{V}$\\
$\mc{W}^{(v)}$  & The list of random walk pairs from view $v \in \mc{V}$\\
$\mbf_u$ & The final embedding of node $u \in \mc{U}$\\ 
$\mbf_u^v$ & The center embedding of node $u \in \mc{U}$ \wrt \, view $v \in \mc{V}$\\
$\tilde\mbf_u^v$ & The context embedding of node $u \in \mc{U}$ \wrt \, view $v \in \mc{V}$\\
\hline
$\theta \in [0, 1]$ & The hyperparameter on parameter sharing in \mtvcon \\
$\gamma \in \mathbb{R}_{\geq 0}$ & The hyperparameter on regularization in \mtvreg \\
$D \in \mathbb{N} $ & The dimension of the embedding space\\
%$N \in \mathbb{N} $ & The number of random walks per view\\
%$L \in \mathbb{N} $ & The length of random walks\\
%$M \in \mathbb{N} $ & The window size for extracting random walk pairs\\
\hline
\end{tabular}
}
\end{table}

%!TEX root = mvn2vec_siam.tex

\section{Preservation and Collaboration in Multi-View Network Embedding}\label{sec::intuition}
In this section, we elaborate on the intuition and presence of \preservation and \collaboration{} -- the two objectives introduced in Section~\ref{sec::introduction}. % and deem important for multi-view network embedding.
We first describe and investigate the motivating phenomena observed in our practice of embedding real-world multi-view networks, and then discuss how they can be explained by the two proposed objectives.

\vpara{Two straightforward approaches for embedding multi-view networks.}
%Most existing network embedding methods~\cite{grover2016node2vec, perozzi2014deepwalk, tang2015line, wang2016structural, perozzi2017don, ou2016asymmetric} are designed for homogeneous networks, where nodes and edges are untyped, while we are interested in studying the problem of embedding multi-view networks.
To extend untyped network embedding algorithms to multi-view networks, two straightforward yet practical approaches exist. 
We refer to these two approaches as the \independent model and the \onespace model.
Specifically, we denote $\mbf^v_u \in \mathbb{R}^{D_{v}}$ the embedding of node $u \in \mc{U}$ achieved by embedding only the view $v \in \mc{V}$ of the multi-view network, where $D_{v}$ is the dimension of the embedding space for network view $v$.
With such notation, the \independent model and the \onespace model are briefly introduced as follows, while further details can be found in Section~\ref{sec::setup}.

\begin{itemize}
\item
\textbf{The independent model}.
Embed each view independently, and then concatenate to derive the final embedding ${\mbf_u}$:
\begin{equation}\label{eg::indep}
\mbf_u = \bigoplus_{v \in \mc{V}} \mbf^v_u \in \mathbb{R}^{D},
\end{equation}
where $D=\sum_{v \in \mc{V}} D_{v}$, and $\bigoplus$ represents concatenation.
In other words, the embedding of each node in the \independent model resides in the direct sum of multiple embedding spaces.
{\textit{This approach preserves the information embodied in each view, but do not allow collaboration across different views in the embedding learning process}}.
\item
\textbf{The one-space model}.
Let the embedding for different views share parameters when learning the final embedding ${\mbf_u}$:
%%%\vspace{-6pt}
\begin{equation}\label{eq::one}
\mbf_u = \mbf^v_u \in \mathbb{R}^{D}, \; \forall v \in \mc{V}. %%%\vspace{-6pt}
\end{equation}
Therefore, the final embedding space correlates with all network views.
{\textit{This approach allows different views to collaborate in learning a unified embedding, but do not preserve information specifically carried by each view}}.
This property of the \onespace model is corroborated by additional experiment in Section~\ref{sec::param-study}.%%%\vspace{-4pt}
\end{itemize}

%In either of the above two approaches, the same treatment to the center embedding is applied to the context embedding when applicable.
Further details on these models can be found in Section~\ref{sec::setup}. 
It should also be noted that the embedding learned by the \onespace model cannot be obtained by linearly combining $\{\mbf^v_u\}_{v \in \mc{V}}$ in the \independent model.
This is because most network embedding models are non-linear.

\begin{table}[tbp]
\centering
\caption{Embedding quality of two real-world multi-view networks using the \independent model and the \onespace model.}\label{tab::exploratory}
%%%\vspace{-9pt}
\resizebox{.35\textwidth}{!}{
\begin{tabular}{| l | c || c | c  |}
 \hline
{Dataset} & {Metric} & {\independent} & {\onespace} \\ \hline 
\multirow{2}{*}{YouTube}  &  ROC-AUC     &\textbf{0.931} &0.914 \\ \cline{2-4}
                                          &  PRC-AUC        &\textbf{0.745} &0.702 \\ \hline
\multirow{2}{*}{Twitter}      & ROC-AUC        &0.724 &\textbf{0.737} \\ \cline{2-4}
                                          & PRC-AUC        &0.447 &\textbf{0.466} \\ \hline
\end{tabular}
}
%%%%\vspace{-6pt}
\end{table}

\vpara{Embedding real-world multi-view networks by straightforward approaches.}
%We highlight the phenomena observed in the practice of embedding real-world multi-view networks using \independent and \onespace.
%In this paper, \independent, and \onespace are implemented on top of a random walk plus skip-gram approach as widely seen in the literature~\cite{grover2016node2vec, perozzi2014deepwalk, perozzi2017don}.
%The experiment setup and results are concisely introduced at this point, while detailed description of the algorithms, datasets, and more comprehensive experiments are deferred to Section~\ref{sec::modeling} and \ref{sec::exp}.
Two networks, YouTube and Twitter, are used in these exploratory experiments with users being nodes on each network.
YouTube has three views representing common videos (cmn-vid), common subscribers (cmn-sub), and common friends (cmn-fnd) shared by each user pair, while Twitter has two views corresponding to replying (reply) and mentioning (mention) among users.
The downstream evaluation task is to infer whether two users are friends, and the results are presented in Table~\ref{tab::exploratory}.

\textit{It can be seen that neither straightforward approach is categorically better than the other.}
In particular, the \independent model consistently outperformed the \onespace model in the YouTube experiment, while the \onespace model outperformed the \independent model in Twitter.
%These exploratory experiments make it clear that neither of the two straightforward approaches is categorically superior to the other.
Furthermore, we interpret the varying performance of the two approaches by the varying extent of needs for modeling \preservation and modeling \collaboration when embedding different networks.
Specifically, recall that the \independent model only captures \preservation, while \onespace only captures \collaboration.
As a result, we speculate if a certain dataset craves for more \preservation than \collaboration, the \independent model will outperform the \onespace model. Otherwise, the \onespace model will win.

\begin{figure}[t]
    \begin{subfigure}[t]{0.2\textwidth}
        \centering
        \includegraphics[width=\textwidth]{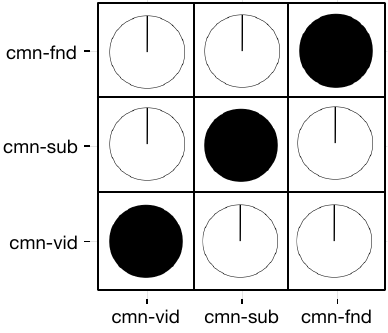}
        \caption{YouTube.}\label{fig::syn-acc-varying-dim}
    \end{subfigure}
    ~
    \hspace*{0.05\textwidth}
    \begin{subfigure}[t]{0.2\textwidth}
        \centering
        \includegraphics[width=\textwidth]{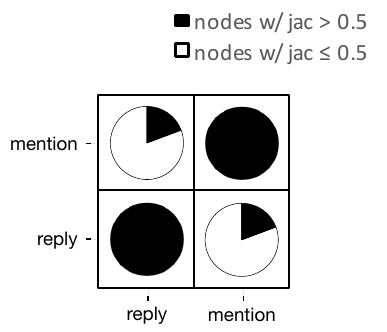}
        \caption{Twitter.}\label{fig::syn-ce-varying-dim}
    \end{subfigure}    
%%%\vspace{-6pt}
\caption{Agreement between information carried by each pair of network views given by a Jaccard coefficient based measurement.}\label{fig::jaccard-exploratory}
\end{figure}

%%To corroborate this speculation, we examine the agreement between different network views.
To corroborate our interpretation of the results, we examine the involved datasets and look into the agreement between information carried by different network views.
We achieve this by a Jaccard coefficient based measurement, where the Jaccard coefficient is a similarity measure with range $[0, 1]$, defined as $J(\mc{S}_1, \mc{S}_2) = |\mc{S}_1 \cap \mc{S}_2| / |\mc{S}_1\cup \mc{S}_2|$ for set $\mc{S}_1$ and set $\mc{S}_2$.
For a pair of network views, a node can be connected to a different set of neighbors in each of the two views.
We calculate the Jaccard coefficient between these two sets of neighbors.
In Figure~\ref{fig::jaccard-exploratory}, we apply this measurement on both datasets and illustrate the proportion of nodes with the Jaccard coefficient greater than $0.5$ for each view pair.

As presented in Figure~\ref{fig::jaccard-exploratory}, little agreement exists between each pair of different views on YouTube.
As a result, it is not surprising that \collaboration among different views is not as needed as \preservation. % in the embedding learning process.
On the other hand, a substantial portion of nodes has Jaccard coefficient greater than $0.5$ over different views on Twitter -- not surprising to see modeling \collaboration brings about more benefits than modeling \preservation.

%\subsection{Further Discussions on Existing Embedding Algorithms}
%We have shown that both \preservation and \collaboration play important roles in embedding multi-view networks.
%However, to the best of our knowledge, no existing embedding methods simultaneously model both objectives.
%%Embedding aims at representing the original network in a Euclidean space. 
%%Therefore, if different view carries different semantic meanings, the learned embedding should reflect it.
%We note that even for methods that do not directly embed different views into the same embedding space, such as~\cite{qu2017attention}, as long as the linear combination is used to find the final embedding from multiple network views, the objectives \preservation will not be modeled.
%This is because the final embedding, in this case, will be a trade-off between the representation among different views, with weights specifying the preference of views to retain information from.
%As a result, information carried by network views with small weights will be lost in the embedding process, which is undesired if an embedding algorithm aims to represent the original network as complete as possible.

%!TEX root = mvn2vec_siam.tex
%\vspace{-6pt}
\section{The \mtv Models}\label{sec::modeling}
In the previous section, \preservation and \collaboration are identified as important objectives for multi-view network embedding.
In the extreme cases, where only \preservation is needed -- each view carries a distinct semantic meaning -- or only \collaboration is needed -- all views carry the same semantic meaning -- simply choosing between \independent and \onespace may be enough to generate satisfactory results.
%In the extreme cases where either \preservation or \collaboration prevails, simply choosing between \independent and \onespace may be enough to generate satisfactory results.
However, it is of more interest to study the scenario where both \preservation and \collaboration co-exist in a multi-view network.
Therefore, we are motivated to (i) study how varied extent of \preservation and \collaboration can impact embedding learning and (ii) explore the feasibility of achieving better embedding by simultaneously modeling both objectives.
To this end, we propose two methods that capture both objectives, while not over-complicating the model or requiring additional supervision.
These two approaches are named \mtvcon and \mtvreg, where \mtv is short for \textbf{m}ulti-\textbf{v}iew \textbf{n}etwork \textbf{to} \textbf{vec}tor, while \textsc{con} and \textsc{reg} stand for constrained and regularized.

As with the notation convention in Section~\ref{sec::intuition}, we denote $\mbf^v_u \in \mathbb{R}^{D_v}$ and $\tilde\mbf^v_u \in \mathbb{R}^{D_v}$ the center and context embedding, respectively, of node $u \in \mc{U}$ for view $v \in \mc{V}$. Further given the network view $v$, \ie, $G^{(v)} = (\mc{U}, \mc{E}^{(v)})$, we use an intra-view loss function to measure how well the current embedding can represent the original network view
%\vspace{-4pt}
\begin{equation}\label{eq::intra-view-loss}
l(\{\mbf^v_u, \tilde{\mbf}^v_u\}_{u \in \mc{U}}| G^{(v)}).
%\vspace{-4pt}
\end{equation}
We defer the detailed definition of this loss function to a later point of this section. 
We let $D_v = D / |\mc{V}| \in \mathbb{N}$ for all $v \in \mc{V}$ out of convenience for model design.
To further incorporate multiple views with the intention to model both \preservation and \collaboration, two approaches are proposed as follows.

\begin{algorithm2e}[t!]
\DontPrintSemicolon
\SetKwInOut{Input}{Input}
\SetKwInOut{Output}{Output}
\Input{ the multi-view network $G = (\mc{U}, \{\mc{E}^{(v)}\}_{v \in \mc{V}})$ and the hyperparameters}
\Output{ the final embedding $\{\mbf_u\}_{u \in \mc{U}}$}
\Begin{
  \For{$v \in \mc{V}$}{
        Sample a list $\mc{W}^{(v)}$ of random walk pairs\;
      }
  Join and shuffle the lists of random walk pairs from all views to form a new random walk pair list $\mc{W}$\;
  \For{each epoch}{
      \tcp{The following for-loop is parallelized}
        \For{$(u, n) \in \mc{W}$}{
            \If{training \mtvcon}{
                Update $\{\mbf^v_u, \tilde{\mbg}^v_u\}_{u \in \mc{U}, v \in \mc{V}}$ with one step descent using gradients in Eq.~\eqref{eq::mtv-con-grad-1}--\eqref{eq::mtv-con-grad-3}
            }       
            \If{training \mtvreg}{
                Update $\{\mbf^v_u, \tilde{\mbf}^v_u\}_{u \in \mc{U}, v \in \mc{V}}$ with one step descent using gradients in Eq.~\eqref{eq::mtv-reg-grad-1}--\eqref{eq::mtv-reg-grad-3}
            }     
        }
      }
  \For{$u \in \mc{U}$}{
   Derive the embedding for node $u$ by $\mbf_u = \bigoplus_{v \in \mc{V}} \mbf^v_u$
  }
}
\caption{\mtvcon and \mtvreg}\label{alg::mtv}
\end{algorithm2e}

\vpara{\mtvcon.}
The \mtvcon model does not enforce further design on the center embedding $\{\mbf^v_u\}_{u \in \mc{U}}$ in the hope of preserving the semantics of each individual view. 
To reflect \collaboration, \mtvcon enforce constraints on the context embedding  $\{\tilde\mbf^v_u\}_{u \in \mc{U}}$ for parameter sharing across different views, so they are required to have the form
%\vspace{-4pt}
\begin{equation}\label{eq::func-phi}
\tilde\mbf^v_u = \phi^v_\theta(\{\tilde\mbg^{v'}_u\}_{v' \in \mc{V}}) \coloneqq (1-\theta) \cdot \tilde\mbg^v_u + \frac{\theta}{|\mc{V}|} \cdot \sum_{v' \in \mc{V}} \tilde\mbg^{v'}_u,
%\vspace{-4pt}
\end{equation}
where $\tilde\mbg^v_u \in \mathbb{R}^{D_v}$ and $\theta \in [0, 1]$ is a hyperparameter controlling the extend to which model parameters are shared. 
The greater the value of $\theta$, the more the model enforces parameter sharing and thereby encouraging more \collaboration across different views.
%This design aims at allowing different views to collaborate by passing information via the shared parameters in the embedding learning process. 
The \mtvcon model solves the following optimization problem
%\vspace{-4pt}
\begin{equation}\label{eq::mtv-con}
\min_{\{\mbf^v_u, \tilde{\mbg}^v_u\}} \sum_{v \in \mc{V}} l(\{\mbf^v_u, \phi^v_\theta(\{\tilde\mbg^{v'}_u\}_{v' \in \mc{V}})\}_{u \in \mc{U}}| G^{(v)}).
%\vspace{-4pt}
\end{equation}
%where $\phi^v_\theta(\{\tilde\mbg^{v'}_u\}_{v' \in \mc{V}})$ is defined in Eq.~\eqref{eq::func-phi}.
After model learning, the final embedding for node $u$ is given by $\mbf_u = \bigoplus_{v \in \mc{V}} \mbf^v_u$.
We note that in the extreme case when $\theta$ is set to be $0$, the model will be identical to the \independent model in Section~\ref{sec::intuition}.
To further distinguish the importance of different views, one can replace $\theta$ in Eq.~\eqref{eq::func-phi} with a view-specific parameter $\theta_v$, we defer the study of which to future work.

\vpara{\mtvreg.}
Instead of setting hard constraints on how parameters are shared across different views, the \mtvreg model regularizes the embedding across different views and solves the following optimization problem
\begin{equation}\label{eq::mtv-reg}
\min_{\{\mbf^v_u, \tilde{\mbf}^v_u\}} \sum_{v \in \mc{V}} l(\{\mbf^v_u, \tilde{\mbf}^v_u\}_{u \in \mc{U}}| G^{(v)}) + \gamma \cdot [\mc{R}^v + \tilde{\mc{R}}^v],
\end{equation}
where  $\gamma \in \mathbb{R}_{\geq 0}$ is a hyperparameter, $\mc{R}^v = \sum_{u \in \mc{U}} \lVert \mbf^v_u - \frac{1}{|\mc{V}|} \sum_{v' \in \mc{V}} \mbf^{v'}_u \rVert^2_2$, $\tilde{\mc{R}}^v = \sum_{u \in \mc{U}} \lVert\tilde\mbf^v_u - \frac{1}{|\mc{V}|} \sum_{v' \in \mc{V}} \tilde\mbf^{v'}_u \rVert^2_2$, and $\norm{\cdot}$ is the $l$-$2$ norm.
This model captures \preservation again by letting $\{\mbf^v_u\}_{u \in \mc{U}}$ and $\{\tilde{\mbf}^v_u\}_{u \in \mc{U}}$ to reside in the embedding subspace specific to view $v \in \mc{V}$, while each of these subspaces are distorted via cross-view regularization to model \collaboration.
Similar to the \mtvcon model, the greater the value of the hyperparameter $\gamma$, the more the \collaboration is encouraged, and the model is identical to the \independent model when $\gamma = 0$.
To distinguish the importance of different views, one can replace $(\sum_{v' \in \mc{V}} \mbf^{v'}_u)/|\mc{V}|$ in the formulation of $\mc{R}^v$ with $({\sum_{v' \in \mc{V}} \lambda_{v'} \cdot \mbf^{v'}_u}) / ({\sum_{v' \in \mc{V}} \lambda_{v'}|}) $, and revise $\tilde{\mc{R}}^v$ similarly.
We also defer this study to future work for simplicity.

\vpara{Intra-view loss function.}
There are many possible approaches to formulate the intra-view loss function in Eq.~\eqref{eq::intra-view-loss}.
In our framework, we adopt the random walk plus skip-gram approach, which is one of the most common in the literature~\cite{grover2016node2vec, perozzi2014deepwalk, perozzi2017don}.
Specifically, for each view $v \in \mc{V}$, multiple rounds of random walks are sampled starting from each node in $G^{(v)} = (\mc{U}, \mc{E}^{(v)})$.
In any random walk, a node $u \in \mc{U}$ and a neighboring node $n \in \mc{U}$ constitute one random walk pair, and a list $\mc{W}^{(v)}$ of random walk pairs can thereby be derived.
We will describe the detailed description of the generation of $\mc{W}^{(v)}$ later in this section.
The intra-view function is then given by
%\vspace{-2pt}
\begin{equation}\label{eq::intra-view-loss-defined}
l(\{\mbf^v_u, \tilde{\mbf}^v_u\}_{u \in \mc{U}}| G^{(v)}) = - \sum_{(u, n) \in \mc{W}^{(v)}} \log p^{(v)}(n | u),
%\vspace{-4pt}
\end{equation}
where $p^{(v)}(n | u) = {\exp\big(\mbf^v_u \cdot \tilde\mbf^v_n\big)}/{\sum_{n' \in \mc{U}} \exp\big(\mbf^v_u \cdot \tilde\mbf^v_{n'}\big)}$.

\vpara{Model inference.}
To optimize the objectives in Eq.~\eqref{eq::mtv-con} and \eqref{eq::mtv-reg}, we opt to asynchronous stochastic gradient descent (ASGD)~\cite{recht2011hogwild} following existing skip-gram--based algorithms~\cite{grover2016node2vec, perozzi2014deepwalk, perozzi2017don, tang2015line, mikolov2013distributed}.
In this regard, $\mc{W}^{(v)}$ from all views are joined and shuffled to form a new list $\mc{W}$ of random walk pairs for all views.
Then each step of ASGD draws one random walk pair from $\mc{W}$ and updates corresponding model parameters with one-step gradient descent.
%Moreover, due to the existence of partition function in Eq.~\eqref{eq::observing-context-given-center}, computing gradients of Eq.~\eqref{eq::mtv-con} and \eqref{eq::mtv-reg} is unaffordable with Eq.~\eqref{eq::intra-view-loss-defined} being their parts.
Negative sampling is adopted as in other skip-gram--based methods~\cite{grover2016node2vec, perozzi2014deepwalk, perozzi2017don, tang2015line, mikolov2013distributed}, which approximates $\log p^{(v)}(n | u)$ in Eq.~\eqref{eq::intra-view-loss-defined} by
$
- \log \sigma(\mbf^v_u \cdot \tilde\mbf^v_n) - \sum_{i=1}^K \expectation_{n'_i \sim P^{(v)}} \log \sigma(-\mbf^v_u \cdot \tilde\mbf^v_{n'_i}),
$
where $\sigma(x)=1/(1+\exp(-x))$ is the sigmoid function, $K$ is the negative sampling rate, $P^{(v)}(u) \propto \left[D_u^{(v)}\right]^{3/4}$ is the noise distribution, and $D_u^{(v)}$ is the number of occurrences of node $u$ in $\mc{W}^{(v)}$~\cite{mikolov2013distributed}.

With negative sampling, the objective function involving one walk pair $(u, n)$ from view $v$ in \mtvcon is
%\vspace{-10pt}
\begin{align*}
\mc{O}_{\textsc{con}} = & \log \sigma(\mbf^v_u \cdot \funcphi{n}) \\
+ & \sum_{i=1}^K \expectation_{n'_i \sim P^{(v)}} \log \sigma(-\mbf^v_u \cdot \funcphi{n'_i}).
%\vspace{-6pt}
\end{align*}

On the other hand, the objective function involving $(u, n)$ from view $v$ in \mtvreg is
%\vspace{-2pt}
\begin{align*}
\mc{O}_{\textsc{reg}} = &  \log \sigma(\mbf^v_u \cdot \tilde{\mbf}^v_n) + \gamma \cdot  (\mc{R}^v_u + \tilde{\mc{R}}^v_n) \\
+ & \sum_{i=1}^K \left[ \expectation_{n'_i \sim P^{(v)}} \log \sigma(-\mbf^v_u \cdot \tilde{\mbf}^v_{n'_i}) + \gamma \cdot  (\mc{R}^v_u + \tilde{\mc{R}}^v_{n'_i}) \right].
%\vspace{-6pt}
\end{align*}
%and $\mc{R}^v_{\hat{u}} = \norm{\mbf^v_{\hat{u}} - \frac{1}{|\mc{V}|} \sum_{v' \in \mc{V}} \mbf^{v'}_{\hat{u}}}^2$, $\tilde{\mc{R}}^v_{\hat{u}} = \norm{\tilde\mbf^v_{\hat{u}} - \frac{1}{|\mc{V}|} \sum_{v' \in \mc{V}} \tilde\mbf^{v'}_{\hat{u}}}^2$.

\begin{figure*}[t]
      \centering
        \includegraphics[width=0.9\textwidth]{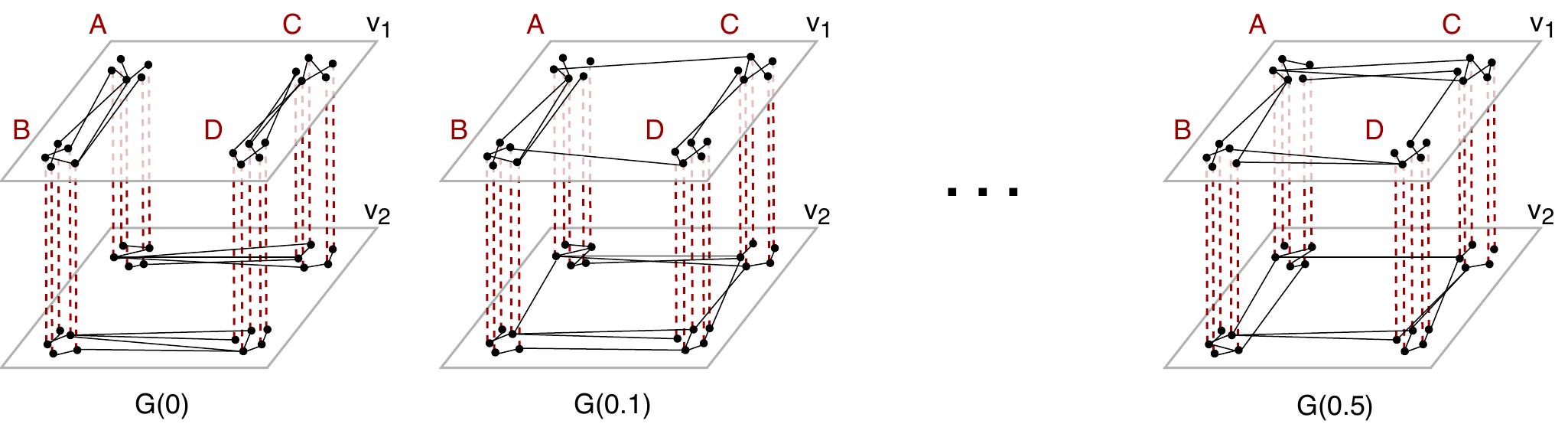}
\caption{An illustration of the series of synthetic networks $G(p)$.}\label{fig::asdf}
\end{figure*}

The gradients of the above two objective function used for ASGD are provided as follows. % provided in the supplementary file. % via an external link due to space limit\footnote{http://yushi2.web.engr.illinois.edu/mvn2vec\_appendix.pdf}. % in the appendix.

\vpara{\mtvcon}:
%\vspace{6pt}
%\begin{footnotesize}
\begin{footnotesize}
\begin{align}\label{eq::mtv-con-grad-1}
\frac{\partial \mc{O}_{\textsc{con}}}{\partial \mbf^v_u} = & \left( 1 - \sigma(\mbf^v_u \cdot \funcphi{n}) \right) \cdot \funcphi{n}  \\ 
- & \sum_{i=1}^K \sigma(\mbf^v_u \cdot \funcphi{n'_i}) \cdot \funcphi{n'_i}, \nonumber
\end{align}
%%\vspace{-6pt}
\begin{equation}\label{eq::mtv-con-grad-2}
\frac{\partial \mc{O}_{\textsc{con}}}{\partial \tilde\mbg^{\hat{v}}_n} =  \left( 1 - \sigma(\mbf^v_u \cdot \funcphi{n}) \right) \cdot \mbf^v_u \cdot
\begin{cases}
\theta + \frac{1 - \theta}{|\mc{V}|},\; & \hat{v} = v,\\
\theta, \; & \hat{v} \neq v,
\end{cases} 
\end{equation}
%%\vspace{-3pt}
\begin{equation}\label{eq::mtv-con-grad-3}
\frac{\partial \mc{O}_{\textsc{con}}}{\partial \tilde\mbg^{\hat{v}}_{n'_i}} = - \sigma(\mbf^v_u \cdot \funcphi{n'_i}) \cdot \mbf^v_u \cdot
\begin{cases}
\theta + \frac{1 - \theta}{|\mc{V}|},\; & \hat{v} = v,\\
\theta, \; & \hat{v} \neq v.
\end{cases}
\end{equation}
\end{footnotesize}
%%\vspace{-6pt}

\vpara{\mtvreg}:
\begin{footnotesize}
\begin{align}\label{eq::mtv-reg-grad-1}
\frac{\partial \mc{O}_{\textsc{reg}}}{\partial \mbf^v_u} = & \left( 1 - \sigma(\mbf^v_u \cdot  \tilde{\mbf}^v_n) \right) \cdot \tilde{\mbf}^v_n - \sum_{i=1}^K  \sigma(\mbf^v_u \cdot  \tilde{\mbf}^v_{n'_i}) \cdot  \tilde{\mbf}^v_{n'_i}    \\ 
+ & 2  \gamma  \Big(K+1\Big)  \Big(1-\frac{1}{|\mc{V}|}\Big) \cdot \Big(\mbf^v_{{u}} - \frac{1}{|\mc{V}|} \sum_{v' \in \mc{V}} \mbf^{v'}_{{u}}\Big), \nonumber
\end{align}
%%\vspace{-6pt}
\begin{align}\label{eq::mtv-reg-grad-2}
\frac{\partial \mc{O}_{\textsc{reg}}}{\partial \tilde\mbf^{{v}}_n} = &  \left( 1 - \sigma(\mbf^v_u \cdot \tilde\mbf^{{v}}_n) \right) \cdot \mbf^v_u 
+ 2  \gamma  \Big(1-\frac{1}{|\mc{V}|}\Big) \cdot \Big(\tilde\mbf^v_{n} - \frac{1}{|\mc{V}|} \sum_{v' \in \mc{V}} \tilde\mbf^{v'}_{n}\Big),
\end{align}
%%\vspace{-6pt}
\begin{align}\label{eq::mtv-reg-grad-3}
\frac{\partial \mc{O}_{\textsc{reg}}}{\partial \tilde\mbf^{{v}}_{n'_i}} = & - \sigma(\mbf^v_u \cdot \tilde\mbf^{{v}}_{n'_i}) \cdot \mbf^v_u 
+ 2  \gamma  \Big(1-\frac{1}{|\mc{V}|}\Big) \cdot \Big(\tilde\mbf^v_{n'_i} - \frac{1}{|\mc{V}|} \sum_{v' \in \mc{V}} \tilde\mbf^{v'}_{n'_i}\Big).
\end{align}
\end{footnotesize}
Note that in implementation, $|V|$ should be the number of views in which $u$ is associated with at least one edge.

%Lastly, we also describe the details on \textbf{random walk pair generation}, provide \textbf{complexity analysis}, and summarize the \mtv algorithms in the supplementary file.

\vpara{Random walk pair generation.}
Without additional supervision, we assume the equal importance of different network views and sample the same number $N \in \mathbb{N}$ of random walks from each view.
To determine $N$, we denote $n^{(v)}$ the number of nodes that are not isolated from the rest of the network in view $v \in \mc{V}$, $n_{\mathit{max}} \coloneqq \max \{ n^{(v)} :\, v \in \mc{V} \}$, and let $N \coloneqq M \cdot n_{\mathit{max}}$, where $M$ is a hyperparameter to be specified.
Given a network view $v \in \mc{V}$, we generate random walk pairs following existing studies~\cite{perozzi2014deepwalk, grover2016node2vec, perozzi2017don}, where the transition probabilities from a node are proportional to the weights of all its outgoing edges.
Specifically, each random walk is of length $L \in \mathbb{N}$, and $\lfloor N / n^{(v)} \rfloor$ or $\lceil N / n^{(v)} \rceil$ random walks are sampled from each non-isolated node in view $v$, yielding a total of $N$ random walks.
For each node in any random walk, this node and any other node within a window of size $B \in \mathbb{N}$ form a random walk pair that is then added to $\mc{W}^{(v)}$.

\vpara{Complexity analysis.}
For every view, random walks are generated independently by existing method~\cite{perozzi2014deepwalk, grover2016node2vec, perozzi2017don}. 
An analysis similar to that in related work~\cite{grover2016node2vec} can show the overall complexity is $\mc{O}\left(\frac{|\mc{V}| \cdot L}{B \cdot (L-B)}\right)$, which is linear to the number of views $|\mc{V}|$.
For model inference, the number of ASGD steps is $|\mc{W}|$, linear to $L$, $B$, $M$ and $|\mc{V}|$.
Each ASGD step computes gradients provided in Eq.~\eqref{eq::mtv-con-grad-1}--\eqref{eq::mtv-reg-grad-3}, which is again linear to $|\mc{V}|$.
Therefore, the overall complexity of model inference is quadratic to the number of views $|\mc{V}|$.

We summarize both the \mtvcon algorithm and the \mtvreg algorithm in Algorithm~\ref{alg::mtv}.

\section{Experiments}\label{sec::exp}
In this section, we further corroborate {{the intuition of \preservation and \collaboration}}, and demonstrate {{the feasibility of simultaneously model these two objectives}}. 
We first perform a case study on a series of synthetic multi-view networks that have varied extent of \preservation and \collaboration. 
Next, we introduce the real-world datasets, baselines, and experiment setting for more comprehensive quantitative evaluations.
Lastly, we analyze the evaluation results and provide further discussion.

\subsection{Case Study -- Varied \preservation and \collaboration on Synthetic Data}\label{sec::synthetic}
To directly study the relative performance of different models on networks with varied extent of \preservation and \collaboration, we design a series of synthetic networks and conduct a multi-class classification task.

We denote each of these synthetic networks by $G(p)$, where $p \in [0, 0.5]$ is referred to as the intrusion probability.
Each $G(p)$ has $4,000$ nodes and $2$ views -- $v_1$ and $v_2$. 
Furthermore, each node is associated to one of the $4$ class labels -- A, B, C, or D -- and each class has exactly $1,000$ nodes.
Before introducing the more general $G(p)$, we first describe the process for generating $G(0)$ as follows:
%\vspace{-3pt}
\begin{enumerate}
\item
generate one random network over all nodes with label A or B, and another over all nodes with label C or D; put all edges in these two random networks into view $v_1$;
%%\vspace{-3pt}
\item
generate one random network over all nodes with label A or C, and another over all nodes with label B or D; put all edges in these two random networks into view $v_2$.
%%\vspace{-3pt}
\end{enumerate}
To generate each of the four aforementioned random networks, we adopt the widely used preferential attachment process with one edge to attach from a new node to existing nodes, where the preferential attachment process is a widely used method for generating networks with power-law degree distribution.
With this design, view $v_1$ carries the information that nodes labeled A or B should be treated differently from nodes labeled C or D, while $v_2$ reflects that nodes labeled A or C are different from nodes labeled B or D.
More generally, $G(p)$ are generated with the following tweak from $G(0)$: when putting an edge into one of the two views, with probability $p$, the edge is put into the other view instead of the view specified in the $G(0)$ generation process.

It is worth noting that greater $p$ favors more \collaboration, while smaller $p$ favors more \preservation. 
In the extreme case where $p=0.5$, only \collaboration is needed in the network embedding process.
This is because every edge has equal probability to fall into view $v_1$ or view $v_2$ of $G(0.5)$, and there is hence no information carried specifically by either view that should be preserved.

\begin{figure}[t]
    \begin{subfigure}[t]{0.225\textwidth}
        \centering
        \includegraphics[width=\textwidth]{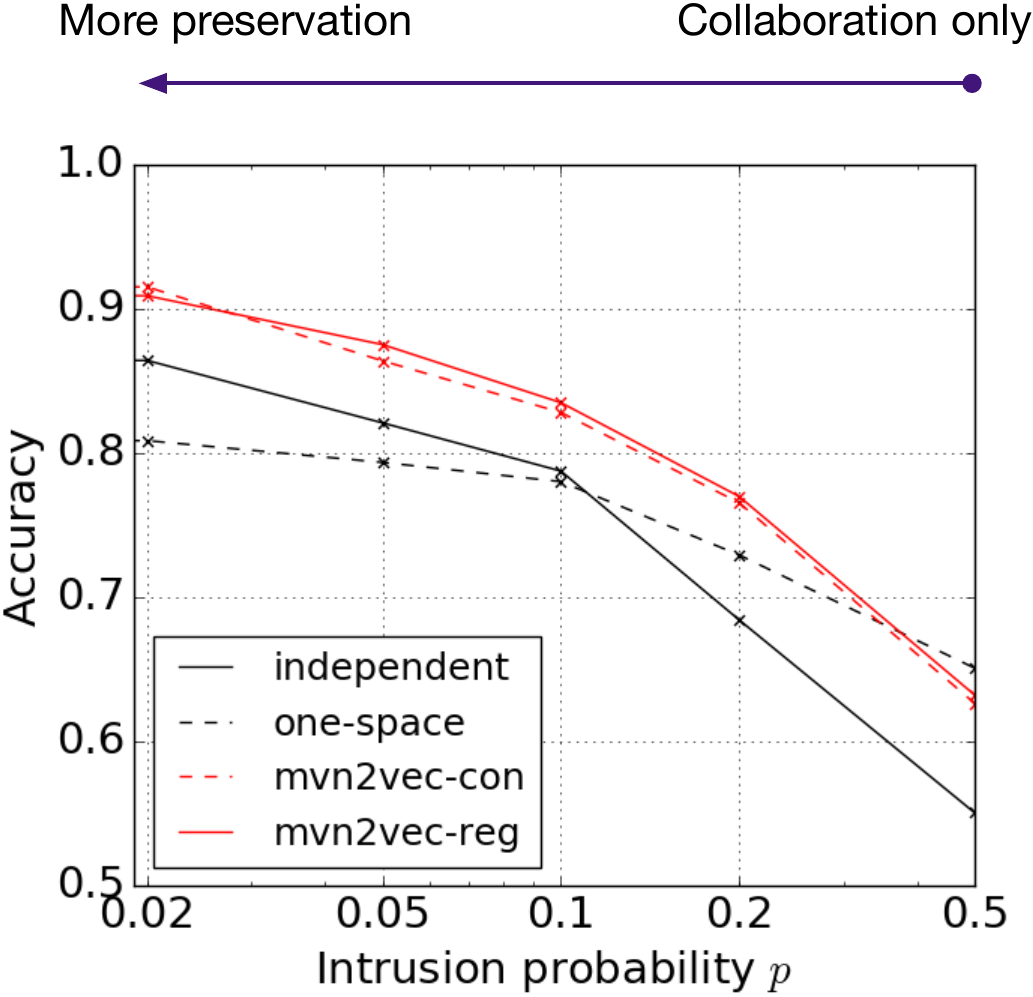}
        %\caption{Accuracy.}\label{fig::syn-acc}
    \end{subfigure}
    ~
    \hspace*{0.01\textwidth}
    \begin{subfigure}[t]{0.225\textwidth}
        \centering
        \includegraphics[width=\textwidth]{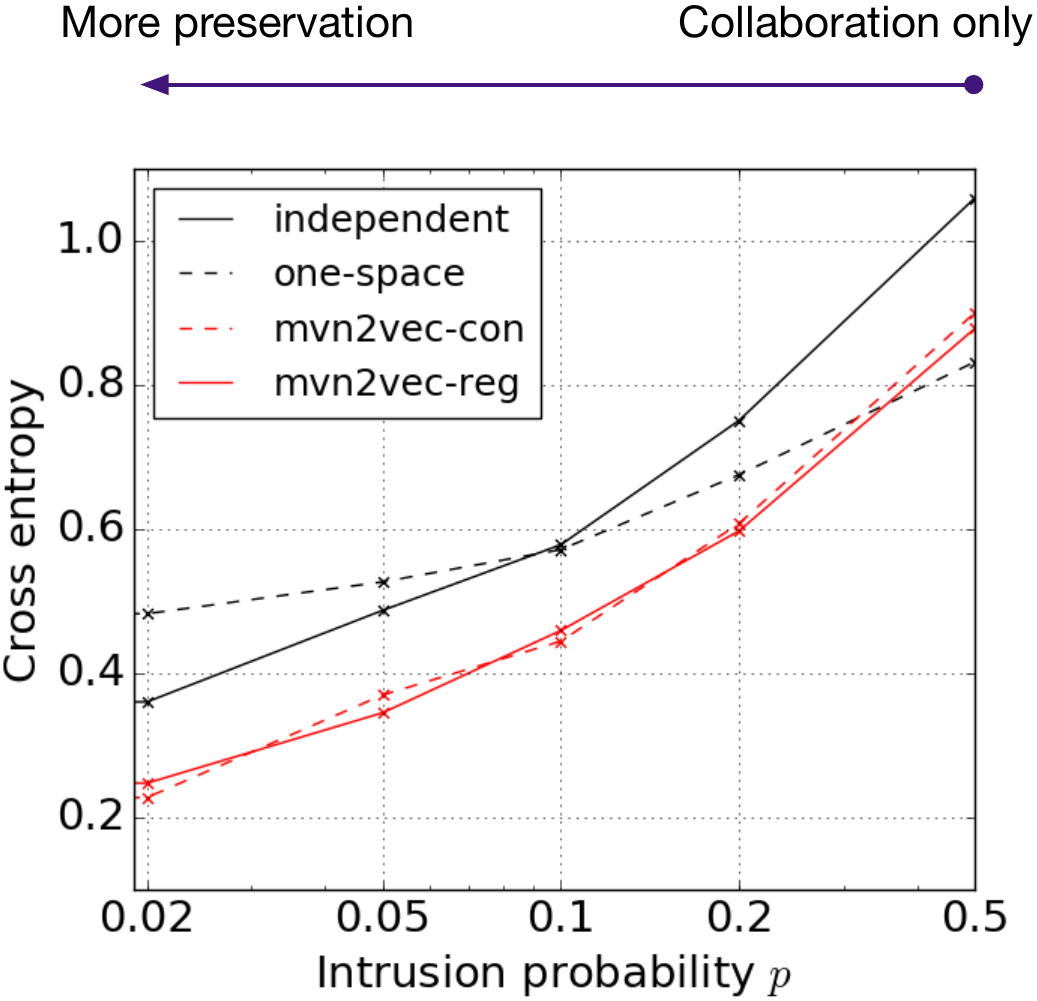}
        %\caption{Cross entropy.}\label{fig::syn-ce}
    \end{subfigure}    
\caption{Classification results under accuracy and cross entropy on synthetic networks $G(p)$ with varied intrusion probability $p$, corresponding to different extent of \preservation and \collaboration.}\label{fig::syn}
\end{figure}

For each $G(p)$, \independent, \onespace, \mtvcon, and \mtvreg are tested.
Atop the embedding learned by each model, we apply logistic regression with cross-entropy to carry out the multi-class evaluation tasks.
Parameters are tuned on a validation dataset sampled from the $4,000$ class labels.
Classification accuracy and cross-entropy on a different test dataset are reported in Figure~\ref{fig::syn}.

\vpara{Observations.} From Figure~\ref{fig::syn}, we make three observations. 
(i) \independent performs better than \onespace in case $p$ is small -- when \preservation is the dominating objective in the network -- and \onespace performs better than \independent in case $p$ is large --  when \collaboration is dominating.
(ii) The two proposed \mtv models perform better than both \independent and \onespace except when $p$ is close to $0.5$, which implies it is indeed possible for \mtv to achieve better performance by simultaneously modeling \preservation and \collaboration.
(iii) When $p$ is close to $0.5$, \onespace performs the best.
This is expected because no \preservation is needed in $G(0.5)$, and any attempts to additionally model \preservation shall not boost, if not impair, the performance.

\subsection{Data Description and Evaluation Tasks}
\begin{figure}[!t]
\noindent\begin{minipage}{.28\textwidth}
\centering
\captionof{table}{Basic statistics for the three real-world networks, where the number of edges specifies the total number of edges from all network views.}\label{tab::data-stats}      
%%\vspace{6pt}
\scalebox{.67}{
\begin{tabular}{| l | c | c | c |}
    \hline
Dataset  & \# views & \# nodes & \# edges  \\ \hline
Snapchat  & 3 & 7,406,859 & 131,729,903 \\  \hline
YouTube  & 3 & 14,900 & 7,977,881 \\  \hline
Twitter & 2 & 116,408 &183,341  \\  \hline
\end{tabular}
}
\end{minipage}
    ~
    \hspace{0.005\textwidth}
\begin{minipage}{.18\textwidth}
 \centering
 \includegraphics[width=\linewidth]{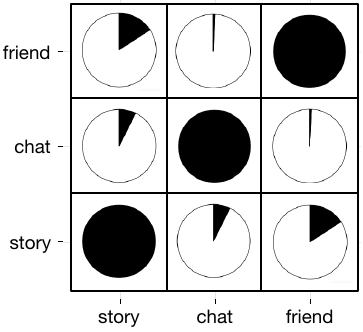}
 %%\vspace{-18pt}
 \captionof{figure}{Agreement between each pair of views for Snapchat.}\label{fig::jaccard-snapchat}  
\end{minipage}
\end{figure}

We perform further quantitative evaluations on three real-world multi-view networks: Snapchat, YouTube, and Twitter.
The key statistics are summarized in Table~\ref{tab::data-stats}, and we describe these datasets as follows.

\vpara{YouTube.}
YouTube is a video-sharing website.
We use the YouTube dataset made publicly available by the Social Computing Data Repository~\cite{zafarani2009social}\footnote{http://socialcomputing.asu.edu/datasets/YouTube}.
From this dataset, a network with three views is constructed, where each node is a core user, and the edges in the three views represent the number of common friends, the number of common subscribers, and the number of common favorite videos, respectively.
Note that the core users are those from which the author of the dataset crawled the data, and their friends can fall out of the scope of the set of core users.
Without user label available for classification, we perform only link prediction task on top of the user embedding.
This task aims at inferring whether two core users are friends, which has also been used for evaluation by existing research~\cite{qu2017attention}.
Each core user forms positive pairs with his or her core friends, and we randomly select $5$ times as many non-friend core users to form negative examples.
%In case a core user has non-friends less than $5$ times of his or her friends, we use all non-friends to generate negative pairs.
Records are split into training, validation, and test sets as in the link prediction task on YouTube.

\vpara{Twitter.}
Twitter is an online news and social networking service.
We use the Twitter dataset made publicly available by the Social Computing Data Repository~\cite{leskovec2014snap}\footnote{https://snap.stanford.edu/data/higgs-twitter.html}.
From this dataset, a network with two views is constructed, where each node is a user, and the edges in the two views represent the number of replies and the number of mentions, respectively.
Again, we evaluate using a link prediction task that infers whether two users are friends as in existing research~\cite{qu2017attention}.
Negative examples generation and training--validation--test split method are the same as in the YouTube dataset.

\vpara{Snapchat.}
Snapchat is a multimedia social networking service. 
On the Snapchat multi-view social network, each node is a user, and the three views correspond to friendship, chatting, and story viewing\footnote{https://support.snapchat.com/en-US/a/view-stories}.
We perform experiments on the sub-network consisting of all users from Los Angeles.
%These users, together with the edges associating them, provide us with a multi-view network.
The data used to construct the network are collected from two consecutive weeks in the Spring of 2017.
Additional data for downstream evaluation tasks are collected from the following week (week 3).
We perform a multi-label classification task, and a link prediction task on top of the same user embedding learned from each network.
For classification, we classify whether or not a user views each of the $10$ most popular discover channels\footnote{https://support.snapchat.com/en-US/a/discover-how-to} according to the user viewing history in week 3, which aims at inferring users' preference and thereby guide product design in content serving.
For each channel, the users who view this channel are labeled positive, and we randomly select $5$ times as many users who do not view this channel as negative examples.
These records are then randomly split into training, validation, and test sets.
This is a multi-label classification problem that aims at inferring users' preference on different discover channels and can therefore guide product design in content serving.
For link prediction, we predict whether two users would view the stories posted by each other in week 3, which aims to estimate the likelihood of story viewing between friends, so as to re-rank stories accordingly.
Negative examples are the users who are friends but do not have story viewing in the same week.
It is worth noting that this definition yields more positive examples than negative examples, which is the cause of a relatively high AUPRC score observed in experiments.
These records are then randomly split into training, validation, and test sets with the constraint that a user appears as the viewer of a record in at most one of the three sets.
This task aims to estimate the likelihood of story viewing between friends so that the application can rank stories accordingly.

We also provide the Jaccard coefficient based measurement on Snapchat in Figure~\ref{fig::jaccard-snapchat}.
It can be seen that the cross-view agreement between each pair of views in the Snapchat network falls in between YouTube and Twitter as presented previously in Section~\ref{fig::jaccard-exploratory}.

For each evaluation task on all three networks, training, validation, and test sets are derived in a shuffle split manner with an $80\%$--$10\%$--$10\%$ ratio. 
We split the data and repeat the experiments $20$ times to compute the mean and its standard error under each metric.
Furthermore, a node is excluded from evaluation if it is isolated from other nodes in at least one of the multiple views.
This processing ensures every node has a valid representation when embedding only one network view.

\subsection{Baselines and Experimental Setup}\label{sec::setup}
In this section, we describe the baselines as well as experimental setup for both embedding learning and downstream evaluation tasks.

\vpara{Baselines.}
Quantitative evaluation results are obtained by applying downstream learner upon embedding learned by a given embedding method.
Therefore, for fair comparisons, we use the same downstream learner in the same evaluation task.
We experiment with both in-house baselines and external baselines from related work.
Since our study aims at understanding the objectives of multi-view network embedding, we build the following in-house embedding methods from the same random work plus skip-gram approach with the same model inference method as described in Section~\ref{sec::modeling}.
%\begin{itemize}
%\item
\textbf{Independent} -- As introduced in Section~\ref{sec::intuition}, the \independent model is equivalent to \mtvcon when $\theta = 0$, and to \mtvreg when $\gamma = 0$.
It preserves the information embodied in each view, but do not allow collaboration across different views in the embedding process.
%\item
\textbf{One-space} -- Also introduced in Section~\ref{sec::intuition}, the \onespace enables different views to collaborate in learning a unified embedding, but do not preserve information specifically carried by each view.
%\item
\textbf{View-merging} -- The \viewmerging model first merges all network views into one unified view, and then learn the embedding of this single unified view.
To comply with the assumed equal importance of all network views, we rescale the weights of edges proportionally in each view to the same total weight. %, so that the total edge weights from all views are the same in the merged network.
This method serves as an alternate approach to \onespace in modeling \collaboration. 
The difference between \viewmerging and \onespace essentially lies in whether or not random walks can cross different views.
We note that just like \onespace, \viewmerging does not model \preservation.
%\item
\textbf{Single-view} -- For each network view, the \singleview model learns embedding from only one view as a sanity check to verify whether introducing more than one view does bring in informative signals in each evaluation task.
In other words, it is identical to embedding only one of the multiple network views using the DeepWalk algorithm~\cite{perozzi2014deepwalk}, or equivalently node2vec with both return parameter and in-out parameter set to $0$~\cite{grover2016node2vec}.
This baseline is used to as a sanity check to verify whether introducing more than one view does bring in informative signals in each evaluation task.
%\end{itemize}

We also experiment with two external baselines that have executable implementation released by the original authors and can be applied to our experiment setting with minimal tweaks.
Note that the primary observations and conclusions made in the experiments would still hold without comparing to these external baselines.
\textbf{MVE}~\cite{qu2017attention} -- An attention-based semi-supervised method for multi-view network embedding. 
The released implementation supports supervision from classification tasks.
For fair comparison, in our unsupervised link prediction experiments, we assign the same dummy class label to $1\%$ randomly selected nodes.
We have observed that the percentage of nodes receiving such dummy supervision does not significantly affect the evaluation results in additional experiments.
\textbf{DMNE}~\cite{ni2018co} -- A deep learning architecture for multi-network embedding, where the multi-network is a more general concept than the multi-view network. 
As a more complex model, the released DMNE implementation takes significantly more time to train on our datasets, and we hence use only the default hyperparameters. 
The authors commented in the original paper that this implementation can be further sped up with additional parallelization.
We experiment with two external baselines on the smaller YouTube and Twitter dataset for scalability reason.

\vpara{Downstream learners.}
For fair comparisons, we apply the same downstream learner onto the features derived from each embedding method.
Specifically, we use the scikit-learn\footnote{http://scikit-learn.org/stable/} implementation of logistic regression with $l$-2 regularization and the SAG solver for both classification and link prediction tasks.
Regularization coefficient in the logistic regression is always tuned to the best on the validation set.
Following existing research~\cite{tang2015line}, each embedding vector is post-processed by projecting onto the unit $l$-2 sphere.
In multi-label classification tasks, the feature is simply the embedding of each node, and an independent logistic regression model is trained for for each label.
In link prediction tasks, features of node pairs are derived by the Hadamard product of the embedding vectors of the two involved node as suggested by previous work~\cite{grover2016node2vec}.

\begin{figure}[t]
    \begin{subfigure}[t]{0.225\textwidth}
        \centering
        \includegraphics[width=\textwidth]{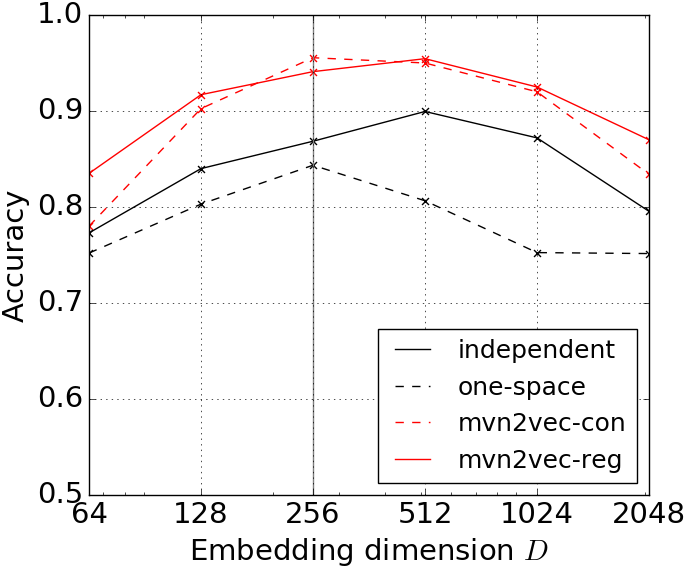}
        %\caption{Accuracy.}\label{fig::syn-acc-varying-dim}
    \end{subfigure}
    ~
    \hspace*{0.01\textwidth}
    \begin{subfigure}[t]{0.225\textwidth}
        \centering
        \includegraphics[width=\textwidth]{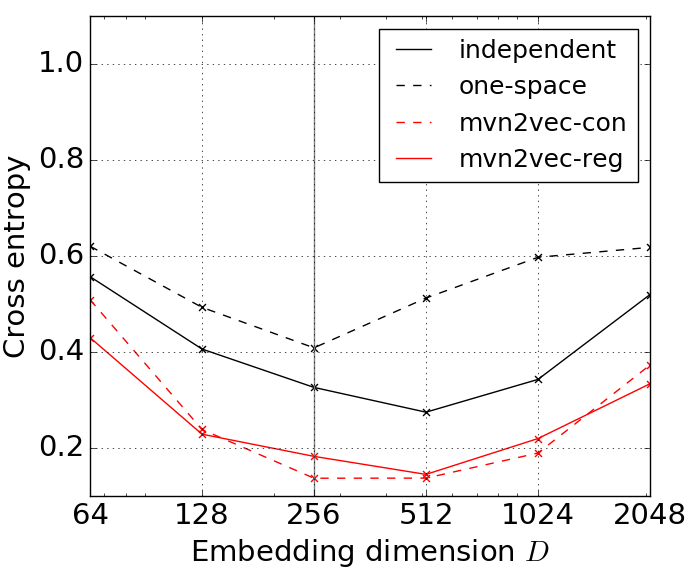}
        %\caption{Cross entropy.}\label{fig::syn-ce-varying-dim}
    \end{subfigure}    
%%%\vspace{-6pt}
\caption{Classification results under accuracy and cross entropy on network $G(0)$ with varied embedding dimension $D$.}\label{fig::syn-varying-dim}
\end{figure}

\vpara{Hyperparameters.}
For \independent, \mtvcon, and \mtvreg, we set embedding space dimension $D = 128 \cdot |\mc{V}|$. 
For \singleview, we set $D = 128$.
For all other methods, we experiment with both $D = 128$ and $D = 128 \cdot |\mc{V}|$, and report the better result.
To generate random walk pairs, we set $L = 20$ and $B = 3$. 
For Snapchat, we set $M=10$ due to its large scale and set $M=50$ for all other datasets.
The negative sampling rate $K$ is set to be $5$ for all models, and each model is trained for $1$ epoch.
In Figure~\ref{fig::syn}, Table~\ref{tab::link-pred-yt-tt}, Table~\ref{tab::quant-snapchat}, and Figure~\ref{fig::syn-varying-dim}, $\theta$ and $\gamma$ in the \mtv models are also tuned on the validation dataset.
The impact of $\theta$ and $\gamma$ on model performance is further discussed in Section~\ref{sec::param-study}.

\vpara{Metrics.}
For link prediction tasks, we use two widely used metrics: the area under the receiver operating characteristic curve (ROC-AUC) and the area under the precision-recall curve (PRC-AUC).
The receiver operating characteristic curve (ROC) is derived from plotting true positive rate against false positive rate as the threshold varies, and the precision-recall curve (PRC) is created by plotting precision against recall as the threshold varies. 
Higher values are preferable for both metrics.
For multi-label classification tasks, we also compute the ROC-AUC and the PRC-AUC for each label and report the mean value averaged across all labels.

%\begin{table*}[t!]
%\centering
%\caption{Mean of classification results with standard error in brackets on the Snapchat multi-view network.}\label{tab::class}
%%%\vspace{-9pt}
%\resizebox{\textwidth}{!}{
%\begin{tabular}{ l | c || c | c | c | c | c || c | c }
%\toprule \hline
%\multirow{2}{*}{Dataset} & \multirow{2}{*}{Metric} & \multicolumn{2}{ c |}{\singleview} & \multirow{2}{*}{\independent} & \multirow{2}{*}{\onespace} & \multirow{2}{*}{\viewmerging} & \multirow{2}{*}{\mtvcon} & \multirow{2}{*}{\mtvreg}  \\ \cline{3-4}
% &  & (worst view) & (best view) &   &   &   &  & \\ \hline
%\multirow{2}{*}{Snapchat}  & ROC-AUC    &0.634 (0.001)&0.667 (0.002)&0.687 (0.001)&0.675 (0.001)&0.672 (0.001)&\textbf{0.693} (0.001)&{0.690} (0.001)\\ \cline{2-9}
%                                          &   PRC-AUC        &0.252 (0.001)&0.274 (0.002)&0.293 (0.002)&0.278 (0.001)&0.279 (0.001)&\textbf{0.298} (0.001)&{0.296} (0.002)\\ \hline
%\bottomrule
%\end{tabular}
%}
%%%%\vspace{-12pt}
%\end{table*}

\begin{table}[]
\centering
\caption{Mean of link prediction results with standard error in brackets on YouTube and Twitter.}\label{tab::link-pred-yt-tt}
%%\vspace{-9pt}
\resizebox{0.49\textwidth}{!}{
\begin{tabular}{ l | c | c | c | c }
\toprule \hline
Dataset            & \multicolumn{2}{c|}{YouTube}                                              & \multicolumn{2}{c}{Twitter}                                              \\ \hline
Metric             & ROC-AUC        & PRC-AUC        & ROC-AUC        & PRC-AUC        \\ \hline \hline
worst \singleview & 0.831 (0.002)                      & 0.515 (0.004)                      & 0.597 (0.001)                      & 0.296 (0.001)                      \\ \hline
best \singleview & 0.904 (0.002)                      & 0.678 (0.004)                      & 0.715 (0.001)                      & 0.428 (0.001)                      \\ \hline
\independent        & \textbf{0.931} (0.001)                      & \textbf{0.745} (0.003)                      & 0.724(0.001)                      & 0.447 (0.001)                      \\ \hline
\onespace           & 0.914 (0.001)                      & 0.702 (0.004)                      & 0.737 (0.001)                      & 0.466 (0.001)                      \\ \hline
\viewmerging        & 0.912 (0.001)                      & 0.699 (0.004)                      & 0.741 (0.001)                      & 0.469 (0.001)                      \\ \hline \hline
MVE~\cite{qu2017attention}                & 0.918 (0.001)                      & 0.714 (0.001)                      & 0.727 (0.001)                      & 0.451 (0.001)                      \\ \hline
DMNE~\cite{ni2018co}               & 0.749 (0.001)                        & 0.311 (0.001)                        & ---                       & ---                       \\ \hline \hline
\mtvcon             & \textbf{0.932} (0.001)                      & \textbf{0.746} (0.001)                      & 0.727 (0.000)                      & 0.453 (0.001)                      \\ \hline
\mtvreg             & \textbf{0.934} (0.001)                      & \textbf{0.754} (0.003)                      & \textbf{0.754} (0.001)                      & \textbf{0.478} (0.001)                      \\ \hline \bottomrule
\end{tabular}
}
\end{table}

\begin{table}[]
\centering
\caption{Mean of link prediction results and of classification results with standard error in brackets on Snapchat.}\label{tab::quant-snapchat}
%%\vspace{-9pt}
\resizebox{0.49\textwidth}{!}{
\begin{tabular}{ l | c | c | c | c }
\toprule \hline
Dataset            & \multicolumn{4}{c}{Snapchat}                                           \\ \hline
Task            & \multicolumn{2}{c|}{Link prediction}                                              & \multicolumn{2}{c}{Classification}                                              \\ \hline
Metric             & ROC-AUC        & PRC-AUC        & ROC-AUC        & PRC-AUC        \\ \hline \hline
worst \singleview & 0.587 (0.001)                      & 0.675 (0.001)                      & 0.634 (0.001)                      & 0.252 (0.001)                      \\ \hline 
best \singleview  & 0.592 (0.001)                      & 0.677 (0.002)                      & 0.667 (0.002)                      & 0.274 (0.002)                      \\ \hline
\independent        & 0.617 (0.001)                      & 0.700 (0.001)                      & 0.687 (0.001)                      & 0.293 (0.002)                      \\ \hline
\onespace           & 0.603 (0.001)                      & 0.688 (0.002)                      & 0.675 (0.001)                      & 0.278 (0.001)                      \\ \hline
\viewmerging        & 0.611 (0.001)                      & 0.693 (0.002)                      & 0.672 (0.001)                      & 0.279 (0.001)                      \\ \hline \hline
\mtvcon             & 0.626 (0.001)                      & 0.709 (0.001)                      & \textbf{0.693} (0.001)                      & \textbf{0.298} (0.001)                      \\ \hline
\mtvreg             & \textbf{0.638} (0.001)                      & \textbf{0.712} (0.002)                      & 0.690 (0.001)                      & 0.296 (0.002)                      \\ \hline \bottomrule
\end{tabular}
}
\end{table}

\begin{figure*}[t]
    \begin{subfigure}[t]{0.235\textwidth}
        \centering
        \includegraphics[width=\textwidth]{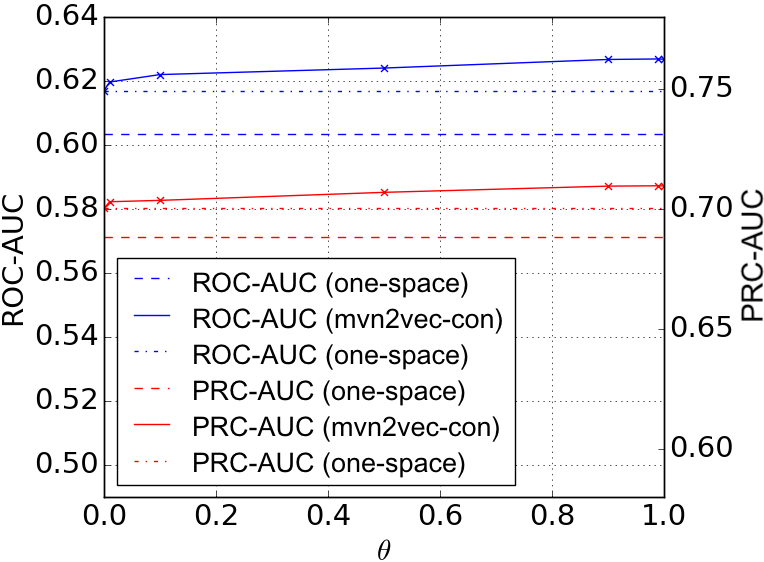}
        \caption{Varying $\theta$ of \mtvcon in Snapchat link prediction.}\label{fig::snapchat-theta-lp}
    \end{subfigure}
    ~
    %\hspace{0.005\textwidth}
    \begin{subfigure}[t]{0.235\textwidth}
        \centering
        \includegraphics[width=\textwidth]{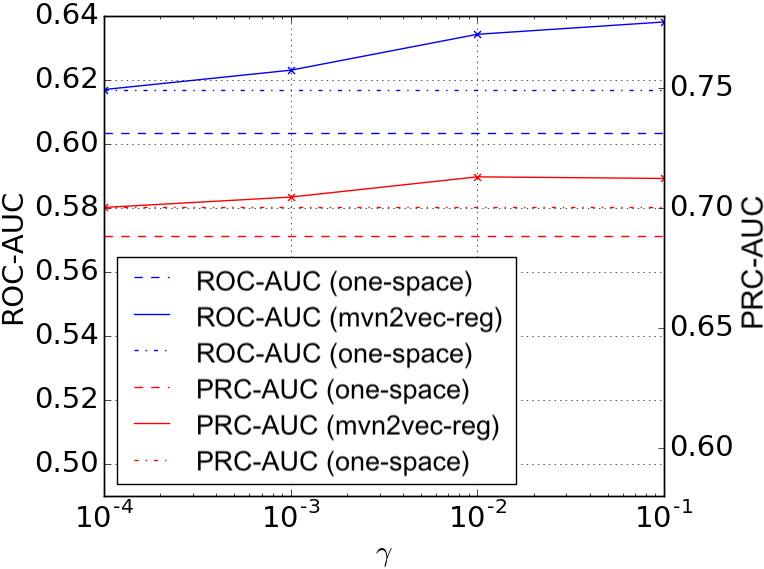}
        \caption{Varying $\gamma$ of \mtvreg in Snapchat link prediction.}\label{fig::snapchat-gamma-lp}
    \end{subfigure}   
    ~
    %\hspace{0.005\textwidth}
    \begin{subfigure}[t]{0.235\textwidth}
        \centering
        \includegraphics[width=\textwidth]{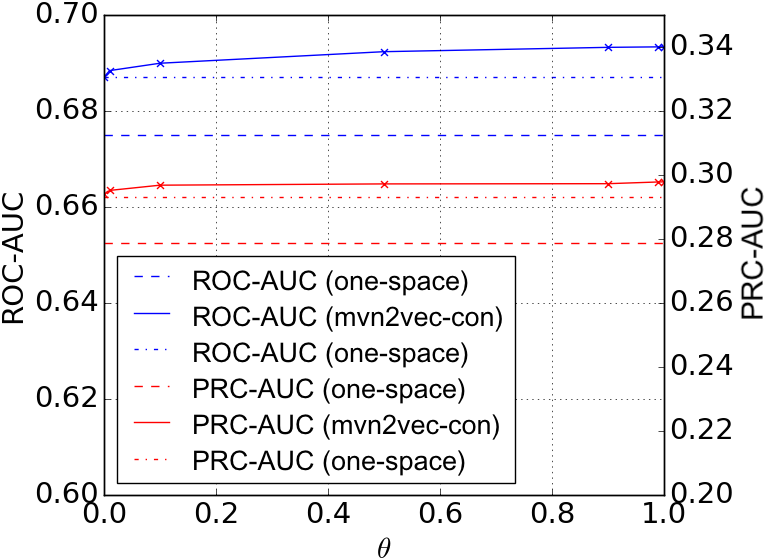}
        \caption{Varying $\theta$ of \mtvcon in Snapchat classification.}\label{fig::snapchat-theta-class}
    \end{subfigure}
    ~
    %\hspace{0.005\textwidth}
    \begin{subfigure}[t]{0.235\textwidth}
        \centering
        \includegraphics[width=\textwidth]{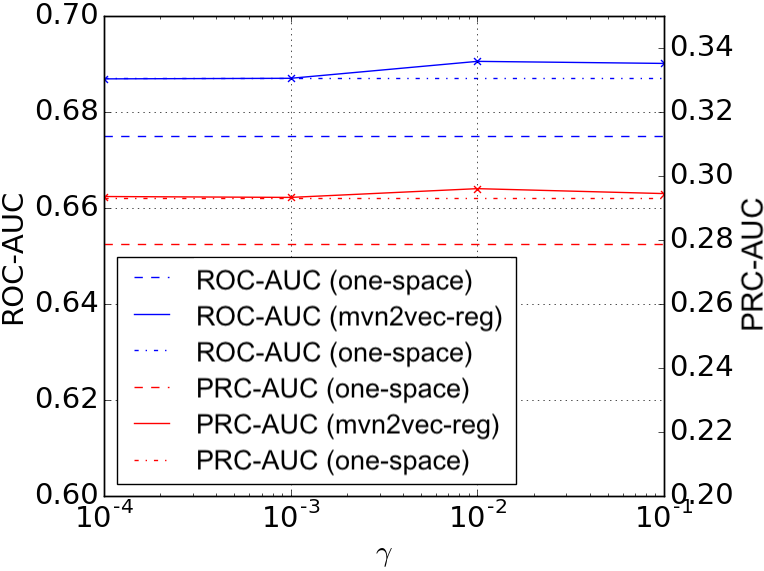}
        \caption{Varying $\gamma$ of \mtvreg in Snapchat classification.}\label{fig::snapchat-gamma-class}
        %%%\vspace{6pt}
    \end{subfigure}   
    
    \begin{subfigure}[t]{0.235\textwidth}
        \centering
        \includegraphics[width=\textwidth]{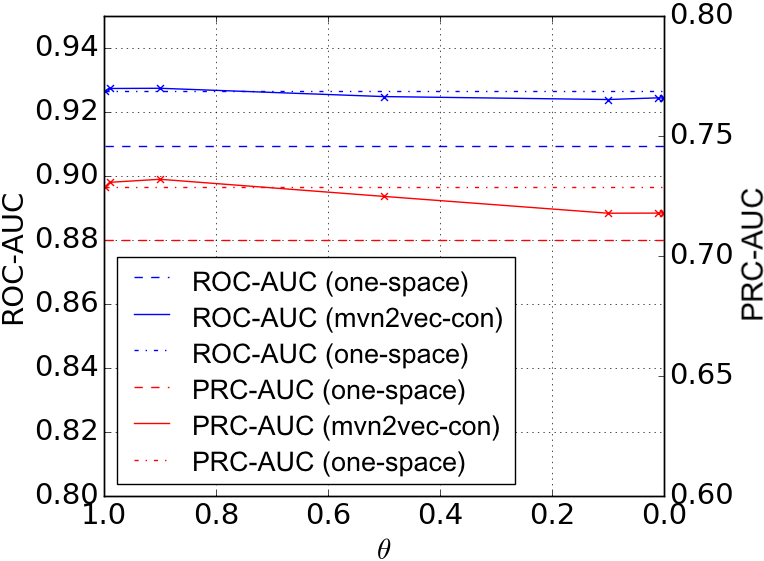}
        \caption{Varying $\theta$ of \mtvcon in YouTube link prediction.}\label{fig::youtube-theta}
    \end{subfigure}
    ~
    %\hspace{0.005\textwidth}
    \begin{subfigure}[t]{0.235\textwidth}
        \centering
        \includegraphics[width=\textwidth]{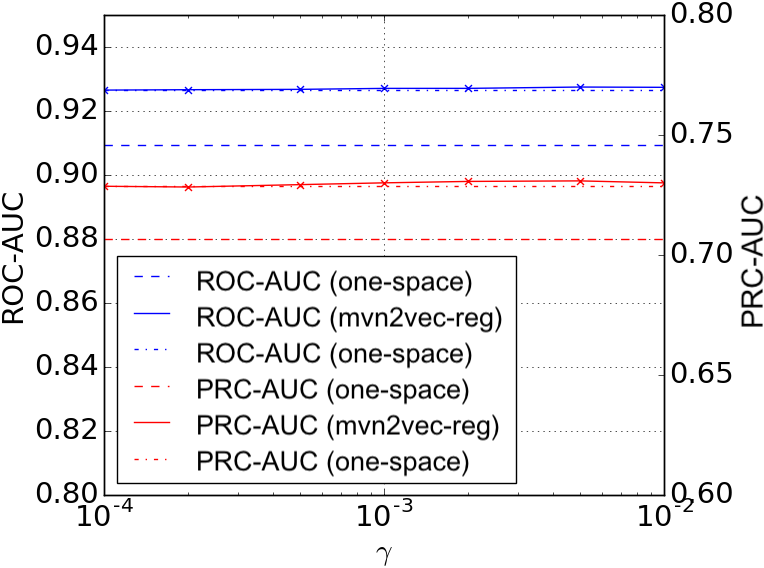}
        \caption{Varying $\gamma$ of \mtvreg in YouTube link prediction.}\label{fig::youtube-gamma}
    \end{subfigure}   
    ~
    %\hspace{0.005\textwidth}
    \begin{subfigure}[t]{0.235\textwidth}
        \centering
        \includegraphics[width=\textwidth]{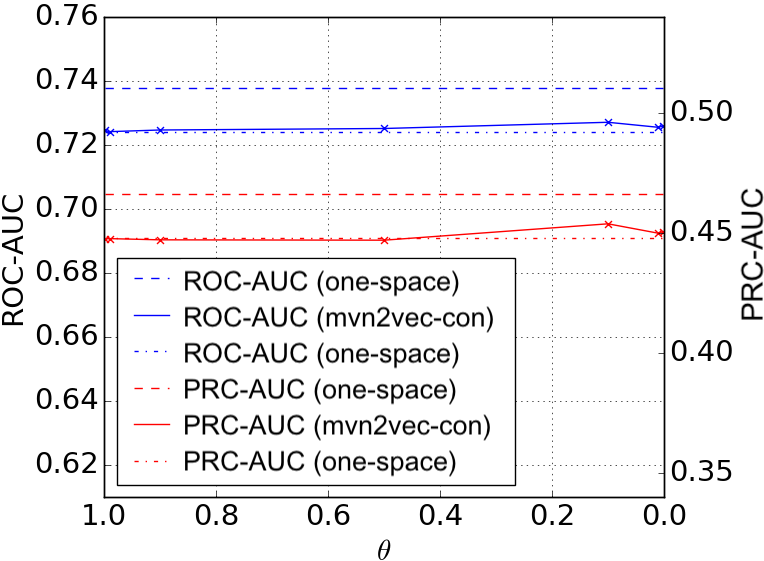}
        \caption{Varying $\theta$ of \mtvcon in Twitter link prediction.}\label{fig::twitter-theta}
    \end{subfigure}
    ~
    %\hspace{0.005\textwidth}
    \begin{subfigure}[t]{0.235\textwidth}
        \centering
        \includegraphics[width=\textwidth]{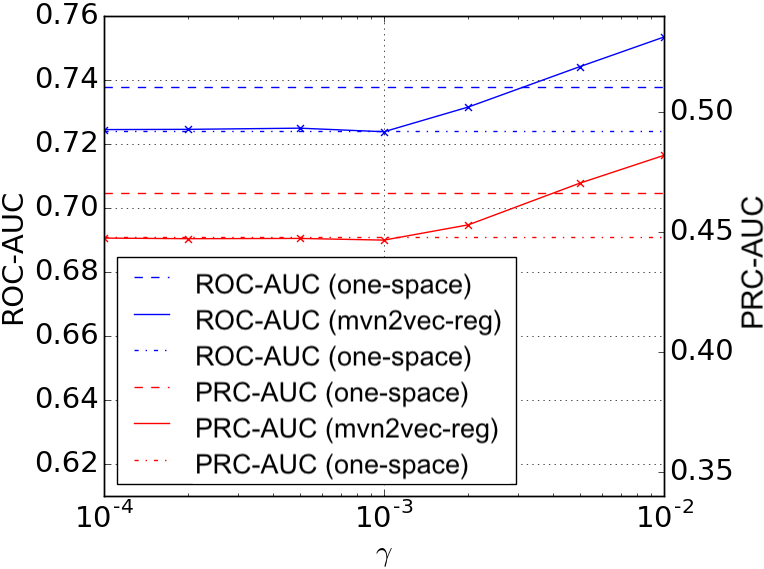}
        \caption{Varying $\gamma$ of \mtvreg in Twitter link prediction.}\label{fig::twitter-gamma}
    \end{subfigure}   
% %%\vspace{-9pt}
 \caption{Performance of the \mtv models under varying hyperparameters regarding \preservation and \collaboration.}\label{fig::varying-theta-gamma}
%%%\vspace{-12pt}
\end{figure*}

\subsection{Quantitative Evaluation Results on Real-World Datasets}\label{sec::quant}

The main quantitative evaluation results are presented in Table~\ref{tab::link-pred-yt-tt} and Table~\ref{tab::quant-snapchat}.
Among the in-house baselines and the proposed \mtv models, {\textit{all models leveraging multiple views outperformed those using only one view}}, which justifies the necessity of using multi-view networks.
Moreover, {\textit{\onespace and \viewmerging had comparable performance}} on each dataset. 
This is an expected outcome because they both only model \collaboration and differ from each other merely in whether random walks are performed across network views.

\vpara{In case the need for either \collaboration or \preservation prevails.}
On YouTube, the proposed \mtv models perform as good but do not significantly exceed the baseline \independent model.
Recall that the need for \preservation in the YouTube network is overwhelmingly dominating as discussed in Section~\ref{sec::intuition}.
As a result, it is not surprising to see that additionally modeling \collaboration does not bring about significant performance boost in such extreme case.
On Twitter, \collaboration plays a more important role than \preservation, as confirmed by the better performance of \onespace than \independent.
Furthermore, \mtvreg achieved better performance than all baselines, while \mtvcon outperformed \independent by further modeling \collaboration, but failed to exceed \onespace. 
This phenomenon can be explained by the fact that $\{\mbf^v_u\}_{v \in \mc{V}}$ in \mtvcon are set to be independent regardless of its hyperparameter $\theta \in [0, 1]$, and \mtvcon's capability of modeling \collaboration is bounded by this design.
%We speculate \mtvcon can potentially outperform \onespace on this dataset if we let $\{\mbf^v_u\}_{v \in \mc{V}}$ to also share parameters via the constraint $\mbf^v_u = (1-\psi) \cdot \mbg^v_u + \frac{\psi}{|\mc{V}|} \cdot \sum_{v' \in \mc{V}} \mbg^{v'}_u$.

\vpara{In case both \collaboration and \preservation are indispensable.}
The Snapchat network used in our experiments lies in between YouTube and Twitter in terms of the need for \preservation and \collaboration.
The proposed two \mtv models both outperformed all baselines under all metrics.
In other words, this experiment result shows the feasibility of gaining performance boost by simultaneously model \preservation and \collaboration without over-complicating the model or adding supervision.

The multi-label classification results on Snapchat are presented in Table~\ref{tab::quant-snapchat}.
As with the previous link prediction results, the two \mtv model both outperformed all baselines under all metrics, with a difference that \mtvcon performed better in this classification task, while \mtvreg outperformed better in the previous link prediction task.
Overall, while \mtvcon and \mtvreg may have different advantages in different tasks, they both outperformed all baselines by simultaneously modeling \preservation and \collaboration on the Snapchat network, where both \preservation and \collaboration co-exist.

\vpara{External baeslines.}
MVE underperformed \mtv models on YouTube.
We interpret this outcome as MVE is a collaboration framework that does not explicitly model \preservation, which has been shown to be needed in the YouTube datasets.
Meanwhile, MVE did not get performance boost from its attention mechanism since the experiment setting forbids it from consuming informative supervision as discussed in Section~\ref{sec::setup}. 
Without additional optimization, the released DMNE implementation did not finish training on the Twitter dataset after two days on a machine with 40 cores of Intel(R) Xeon(R) CPU E5-2680 v2 @ 2.80GHz.
The comparatively worse performance of DMNE on YouTube should partially attribute to the use of default hyperparameters as described in Section~\ref{sec::setup}.
Another possible explanation is that DMNE is designed for the more general multi-networks, not optimized for multi-view networks.

These observations in combine demonstrated the {{feasibility of gaining performance boost by simultaneously modeling \preservation and \collaboration without over-complicating the model or requiring additional supervision}}.

\subsection{Hyperparameter Study}\label{sec::param-study}
\vpara{Impact of embedding dimension.} 
To rule out the possibility that \onespace could actually preserve the view-specific information as long as the embedding dimension were set to be large enough, we 
study the impact of embedding dimension.
Particularly, we carry out the multi-class classification task on $G(0)$ under varied embedding dimensions.
Note that $G(0)$ is used in this experiment because it has the need for modeling \preservation as discussed in Section~\ref{sec::synthetic}.
As presented in Figure~\ref{fig::syn-varying-dim}, \onespace achieves its best performance at $D=256$, which is worse than 
\independent at $D=256$, let alone the best performance of \independent at $D=512$.
Therefore, one cannot expect \onespace to preserve the information carried by different views by employing embedding space with large enough dimension.

Besides, all four models achieve their best performance with $D$ around 256$\sim$512. 
Particularly, \onespace uses the smallest embedding dimension to reach peak performance. 
This is expected since \onespace does not divide the embedding space to suit multiple views and have more freedom in exploiting an embedding space with given dimension.

\vpara{Impact of $\theta$ for \mtvcon and $\gamma$ for \mtvreg.}
We also study the \textbf{impact of $\theta$ for \mtvcon and $\gamma$ for \mtvreg}, which corroborated the observations in Section~\ref{sec::quant}.
With results presented in Figure~\ref{fig::varying-theta-gamma}, we first focus on the Snapchat network.
Starting from $\gamma=0$, where only \preservation was modeled, \mtvreg performed progressively better as more \collaboration kicked in by increasing $\gamma$.
The peak performance was reached between $0.01$ and $0.1$.
On the other hand, the performance of \mtvcon improved as $\theta$ grew. 
Recall that even in case $\theta=1$, \mtvcon still have $\{\mbf^v_u\}_{v \in \mc{V}}$ independent in each view.
This prevented \mtvcon from promoting more \collaboration.

On YouTube, the \mtv models did not significantly outperform \independent no matter how $\theta$ and $\gamma$ varied due to the dominant need for \preservation as discussed in Section~\ref{sec::intuition} and \ref{sec::quant}.

On Twitter, \mtvreg outperformed \onespace when $\gamma$ was large, while \mtvcon could not beat \onespace for reasons discussed in Section~\ref{sec::quant}.
This also echoed \mtvcon's performance on Snapchat as discussed in the first paragraph of this section.

%!TEX root = mvn2vec_siam.tex

\section{Conclusion and Future Work}
We identified and studied the objectives that are specific and important to multi-view network embedding, \ie{} \preservation and \collaboration.
We then explored the feasibility of better embedding by simultaneously modeling both objectives, and proposed two multi-view network embedding methods.
Experiments with various downstream tasks were conducted on a series of synthetic networks and three real-world multi-view networks from distinct sources, including two public datasets and a large-scale internal Snapchat dataset.
The results corroborated the validity and importance of \preservation and \collaboration as two optimization objectives, and demonstrated the effectiveness of the proposed \mtv methods.

Knowing the existence of the identified objectives, future works include modeling different extent of \preservation and \collaboration for different pairs of views in multi-view embedding.
It is also rewarding to explore supervised methods for task-specific multi-view network embedding that jointly model \preservation and \collaboration.% as optimization objectives.

\vpara{Acknowledgments.}
This work was sponsored in part by U.S. Army Research Lab. under Cooperative Agreement No. W911NF-09-2-0053 (NSCTA), DARPA under Agreement No. W911NF-17-C-0099, National Science Foundation IIS 16-18481, IIS 17-04532, and IIS-17-41317, DTRA HDTRA11810026, and grant 1U54GM114838 awarded by NIGMS through funds provided by the trans-NIH Big Data to Knowledge (BD2K) initiative (www.bd2k.nih.gov). Any opinions, findings, and conclusions or recommendations expressed in this document are those of the author(s) and should not be interpreted as the views of any U.S. Government. The U.S. Government is authorized to reproduce and distribute reprints for Government purposes notwithstanding any copyright notation hereon.

\bibliographystyle{IEEEtran}
\bibliography{yushi} 

\newpage
\end{document}